\documentclass[5p,times]{elsarticle}
\usepackage{lineno,hyperref}
\usepackage{mathptmx}  
\usepackage{amsmath}
\usepackage{amsfonts}
\usepackage{amssymb}
\usepackage{graphicx}
\usepackage{subfig}
\usepackage{soul}
\usepackage[dvipsnames]{xcolor}
\usepackage{romannum}
\usepackage[normalem]{ulem}
\usepackage{enumerate}
\usepackage{lscape}

\usepackage[normalem]{ulem}

\usepackage{tikz}
\usepackage[english]{babel}
\usepackage{subfig}
\usepackage{paralist}
\usepackage{multirow}
\usepackage{booktabs}
\usepackage{caption}
\usepackage{framed}
\usepackage{listings}
\usepackage[ruled,linesnumbered]{algorithm2e}
\usepackage{xcolor, colortbl}    
\usepackage{soul}
\usepackage{pifont}
\usepackage{balance}
\hypersetup{
    colorlinks,
    linkcolor={black!50!black},
    citecolor={blue!50!black},
    urlcolor={blue!80!black}
}

\usepackage{tikz}
\usepackage[english]{babel}
\usepackage{hyperref}
\usepackage{url}
\usepackage{subfig}
\usepackage{paralist}
\usepackage{multirow}
\usepackage{booktabs}
\usepackage{caption}
\usepackage{framed}
\usepackage{listings}
\usepackage{algorithmic,algorithm2e,float}
\usepackage{soul}
\usepackage{pifont}
\usepackage{balance}
\hypersetup{
	colorlinks,
	linkcolor={black!50!black},
	citecolor={blue!50!black},
	urlcolor={blue!80!black}
}
\SetKwInput{KwInput}{Input}                
\SetKwInput{KwOutput}{Output}              

\date{}

\def\BibTeX{{\rm B\kern-.05em{\sc i\kern-.025em b}\kern-.08em
    T\kern-.1667em\lower.7ex\hbox{E}\kern-.125emX}}
\begin{document}

\begin{frontmatter}

\title{A Giant-Step Baby-Step Classifier For Scalable and Real-Time Anomaly Detection In Industrial Control Systems and Water Treatment Systems}

\author{ Sarad Venugopalan, Sridhar Adepu}
\address{Swansea University, Swansea, United Kingdom}




\begin{abstract}
The continuous monitoring of the interactions between cyber-physical components of any industrial control system (ICS) is required to secure automation of the system controls, and to guarantee plant processes are fail-safe and remain in an acceptably safe state. Safety is achieved by managing actuation (where electric signals are used to trigger physical movement), dependent on corresponding sensor readings; used as ground truth in decision making. 	Timely detection of anomalies (attacks, faults and unascertained states) in ICSs is crucial for the safe running of a plant, the safety of its personnel, and for the safe provision of any services provided. 
We propose an anomaly detection method that involves accurate linearization of the non-linear forms arising from sensor-actuator(s) relationships, primarily because solving linear models is easier and well understood. We accomplish this by using a well-known water treatment testbed as a use case. Our experiments show millisecond time response to detect anomalies, all of which are explainable and traceable; this simultaneous coupling of detection speed and explainability has not been achieved by other state of the art Artificial Intelligence (AI)/ Machine Learning (ML) models with eXplainable AI (XAI) used for the same purpose.
Our methods explainability enables us to pin-point the sensor(s) and the actuation state(s) for which the anomaly was detected.
The proposed algorithm showed an accuracy of 97.72\% by flagging deviations within safe operation limits as non-anomalous; indicative that slower detectors with highest detection resolution is unnecessary, for systems whose safety boundaries provide leeway within safety limits.
\end{abstract}

\begin{keyword}
\texttt{Anomaly Detection, Explainable, Real Time, Scalable, Energy Efficient.}
\end{keyword}
\end{frontmatter}

	\section{Introduction}
	\label{sec:intro}
	\footnote{This work was carried out when the authors were employed at University of Bristol in the United Kingdom. Sarad Venugopalan was at Swansea University.}
	Physical/mechanical automation components are more recently becoming digitized and analogue systems are also being developed with digital interfaces to enable integration with digital systems. 
	Industries such as manufacturing (including assembly, quality check and distribution) and critical infrastructure (CI) - including the water sector (water and wastewater treatment, and distribution), energy sector (generation, transmission and distribution), and transportation sector, play an important role in global society. The provision of high-quality CI services in particular should be noted as significant for public health, well-being and safety.
	The integration and embedding of computational processing units expose the ICS used in industries and CI to an
	expanded attack surface, making it vulnerable to cyber-attacks~\cite{Jaikaran2023}. 
	For example,  malware was found in the manufacturing sector focused on a steel plant~\cite{steelplant2015}. This attack led to an unregulated furnace and the resulting dangerous infrastructure could not be shut down as normal, leading to physical damage to the plant~\cite{Sans2014}.  Recent power sector-focused malware such as 
	FrostyGoop~\cite{FrostyGoop2024} were able to infiltrate CI and disrupt services.  
	Disruption and damage have been recorded in the water sector due to attacks on CPS at the Maroochy Shire plant in Australia~\cite{MaroochyShire2008}, multiple plants in Texas, USA~\cite{Texas2024}, Mayo in Ireland~\cite{Mayo2023}, and a mitigated attack at the Oldsmar plant in USA~\cite{Oldsmar2021}.
	The spate of real-world attacks on manufacturing and CI operational technology (OT), and their consequences reiterate the value in timely detection of these attacks and subsequent actions to bring these systems into a safe and non-compromised state. 
	
	In this work, we use the Secure Water Treatment (SWaT)~\cite{Goh2016ADT} testbed as a use case in ICS anomaly detection.
	We use system-specific information to improve anomaly detection, as opposed to other models that employ specialized solutions on a subset of available information~\cite{ECOD, SVDD, Mathuros2024}, or prefer to use a generic data-driven approach~\cite{Feng2019}.
	Specifically, we rely on \textit{i)}  the real-time data logs to retrieve state information from plant sensors and actuators, \textit{ii)} the PLC control logic sheet to determine sensor boundary conditions and corresponding actuation steps (if any), and \textit{iii)} infer sensor-actuator(s) relationships from the plant system architecture. 
	
	Going over the complete actuation state space is a case of solving problems in exponential time (see Section~\ref{ssec:lincomplexity}).
	We employ a dimensionality reduction technique to reduce the time complexity by considering only the nearest-neighbour actuators related to a given sensor.
	This is followed by an accurate linearization of a reduced number of actuator(s)-sensor relationships.
	We note that the water treatment plant (see Fig.~\ref{fig:swat}), has limited nonlinear sensor-actuator(s) relationships. 
	This is attributed to the (mostly) serial nature of numerous industrial processes, including those involving assembly lines in a factory setting and a water treatment plant.  
	Once the sensor-actuator(s) relationships are linearized, it is trivial to determine the bounds required to detect anomalies in the ICS. The advantage of this approach is that it typically allows detection and explanation within a millisecond (see Section~\ref{sec:experiments}). It saves valuable time with fast detection and explanation,  helping with faster diagnostics, mitigation, or recovery. 
	The use of per sensor anomaly detection boundaries makes our solution scalable to a number of mostly serial industrial processes (see Section~\ref{sssec:sensorindependence})  
	The novelty being that our solution is fit for use in near real-time and resource-constrained decision-making control systems, and at the same time provides accurate explanation and traceability.
	We make the following main contributions. 
	\begin{compactenum}

		\item [\textit{i}.)] We present our anomaly detector in Section~\ref{sec:solution}.
		
		\item [\textit{ii}.)] An extended solution is proposed in Section~\ref{sec:extended} to detect anomalies that stand out over a  longer period of time.
		
		\item [\textit{iii}.)]  We compare our solution results with AI/ML models for anomaly detection and explainable AI  (see Section~\ref{sec:experiments}).
		
		\item [\textit{iv}.)] We discuss how explainability~\cite{Ali2023} is incorporated into our solution (see Section~\ref{ssec:Explainablity}).
		Reducing operator workload while maintaining safety is discussed in Section~\ref{ssec:operatorworkload}.
		
		\item [\textit{v}.)] We discuss the real-time and scalable nature of the proposed solution (see Section~\ref{ssec:scalable}) making it a candidate for use in other constrained devices, edge computing devices and legacy systems with lower computational power (see Section~\ref{ssec:Usability}).
		
	\end{compactenum}
	
	Our contributions highlight the benefits of using deterministic and unembellished solutions to meeting stakeholder and security requirements. Our solution provides scalable and real-time detection, and guaranteed explainability. It is made possible using system specific knowledge. This makes it a worthy competitor w.r.t. other deep learning AI algorithms (used for the same purpose), with complex explainability addons.
    The paper is organised as follows.
	Section~\ref{sec:background} provides the problem context and discusses the threat model. Section~\ref{sec:solution} presents the detector solution. The solution is extended in Section~\ref{sec:extended}.
	Experiments are carried out and results presented in Section~\ref{sec:experiments}.
	The work is discussed in Section~\ref{sec:discussion}.
	The related work is provided in Section~\ref{sec:litreview}.
	The conclusions are drawn, and future work is proposed in Section~\ref{sec:conclusions}.

	\begin{figure}
		\centering
		\includegraphics[width=1.0\linewidth]{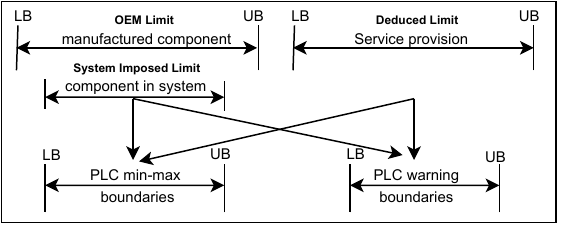}
		\caption{\small Safe state determination using sensors. Boundaries are fed into a programmable logic controller. Lower bound (LB) and upper bound (UB).}
		\label{fig:boundaries}
	\end{figure}

	\section{Problem Context and Threat Model}
	\label{sec:background}

	\subsection{Sensor readings and boundary conditions}
	\label{ssec:boundary}
	In context, boundaries refer to safety limits for plant operations. These limits (see Fig.~\ref{fig:boundaries}) apply to \textit{(i})  transport components' operational characteristics (such as pump speed and pipe pressure), and (\textit{ii}) service provisions (such as the chemicals required to maintain the hydrogen ion concentration (pH) and oxidation-reduction potential (ORP) for water, in the required safe range for consumption). 
	The distinction between transport characteristics and service provisioning is made based on requirement. For example, pipes and pumps may be used to safely transport a variety of fluids, whereas service provisions for water treatment implies that provisioning used is specific for water quality and consumption safety.
	A service is most often provided using a conglomeration of different transport components and service provisions. 
	
	Transportation components from reputable manufacturers are supplied with detailed specifications, defining the acceptable safe operational range and conditions of use. When used as part of a system, a new appropriate system-imposed limit may apply depending on the system design and other operational bottlenecks.
	For example, a pipe may be Original Equipment Manufacturer (OEM) safety rated for a pressure of 0-5 Bar. However, downstream constraints may only allow a pressure of 0-3 Bar. In such cases, a system-imposed limit is put in place.
	With respect to service provisioning boundaries, they are deduced from extensive knowledge and tests to determine the safe range for consumption.
	For example, a water treatment process must ensure that adding acid and/or base chemicals do not significantly result in the pH of water moving away from 7 (neutral). 
	
	The OEM/system imposed limits for transport components and deduced limits for service provisioning parts are fed into one or more controllers, such as a PLC (see Fig.~\ref{fig:boundaries}). The min-to-max operational range values on the PLC assists with taking action and (possibly) alerting an operator, when the safety threshold is breached (see Section~\ref{ssec:operatorworkload}). Additionally, the PLCs may also include logic to act, even before a breach, by providing a warning buffer. This may be achieved by setting tighter boundary conditions. Traditionally, at least two layers of safety are provided, the lower boundary is the warning boundary which alerts operators to take action, the upper boundary is the min-max boundary, and this should never be passed.
	In some systems if the min-max boundary is passed the personnel must evacuate the site and services are put on hold until safe conditions are assured. 
	
	\begin{figure}
		\centering
		\includegraphics[width=1.0\linewidth]{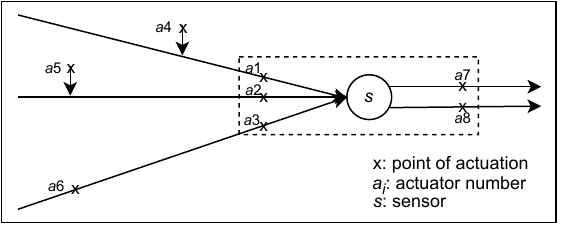}
		\caption{\small A sensor and dependent series-parallel actuators in a physical process. Its nearest neighbor actuator(s) are shown inside the dotted rectangle. }
		\label{fig:nonlinear}
	\end{figure}
	
	\subsection{Reducing time complexity: core idea}
	\label{ssec:lincomplexity}
	In this section, we explain the core idea on which we base our solution.
	The set of boundaries [LB, UB] in anomaly detection, used as safe range for operations may not always be trivial to optimally determine. 
	Taking into consideration all actuators together causes a time complexity blow up, resulting in a longer time to solve the problem optimally. 
	For example, let each actuator have $k$ actuation sequences (such as open/transition/close; here $|k|=3$), and let there be $n$ such actuators, in relation with a given sensor $s$. Hence, there are a total of $k^n$ actuation states, in relation with sensor $s$. 
	In Fig.~\ref{fig:nonlinear}, the readings from a sensor $s$ may change due to actuation in plant processes, at points marked X. The actuators at the input side are named $a_1-a_6$, and the output side are $a_7-a_8$.   
	Now, if we can reduce the state space for sensor relations to its immediate neighbour actuators  ($a_1-a_3$ and $a_7-a_8$; see dotted rectangle in Fig.~\ref{fig:nonlinear}), we reduce our solution seek time. Therefore, we employ a dimensionality reduction of the actuator(s) in relation with sensor $s$.
	
\begin{figure}
		\centering
		\includegraphics[width=0.75\columnwidth]{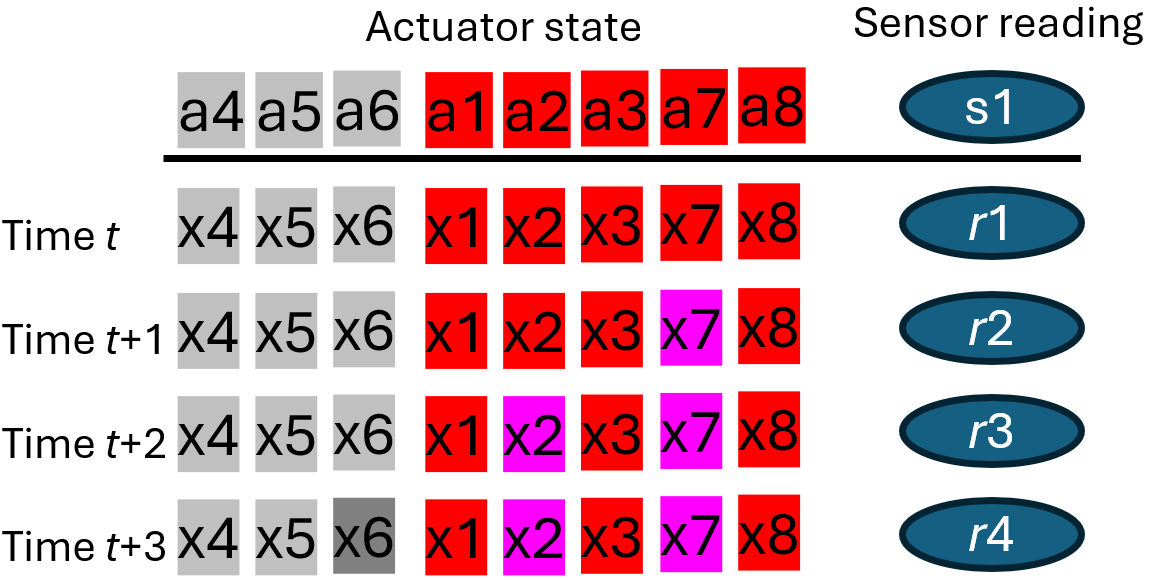}
		\caption{\small The actuators a4-a6 are for non-nearest neighbors w.r.t. Fig.~\ref{fig:nonlinear}. a1-a3 and a7-a8 are nearest neighbors. The $x_i$'s are their actuation state. Sensor $s = s_1$ is considered here. At times $t+1$, $t+2$, $t+3$, the actuation state of a7, a2 and a6, namely, $x7$, $x2$, and $x6$ changed. For brevity  only binary actuation is shown. Even when a4-a6 are ignored, the sensor reading and its overall boundaries will remains unchanged. However, ignoring the non-nearest neighbors might result in a loss in detection resolution. This further depends on whether those actuators are capable of affecting the sensor reading. Any included actuator whose states does not affect the sensor reading may be considered as a do not care actuator. In the case the sensor-actuator(s) relationship is available from the design document or an expert is able to deduce those relationships, it may be possible to further refine the actuators related to the sensor --- to minimise resolution loss.}
		\label{fig:truthtable}
	\end{figure}

	Let there be normal and attack plant operation datasets for the sensor and actuator states logged during plant operations, under normal and attack conditions, respectively.
	A normal training dataset comprises of all seen (normal) state relations embedded into it. 
	We implicitly rely on all actuator relations $f(a_1,\cdots,a_8$) w.r.t. the sensor, but explicitly consider the actuation sequences for $a_1-a_3$, for input side linearization. 
	The excluded actuators $a_4-a_6$ are indirectly responsible for the change in measurement readings at sensor $s$. However, these relations are captured at the nearest neighbour (nn-) actuators\footnote{Henceforth, it is referred to as  nn-actuators. It is common to place the actuator immediately preceding the related sensor. If not, we yet consider that directly related actuator as a nearest neighbour --- to improve detector resolution, when that information is available. Resolution is further enhanced by picking non-neighbour PLC imposed actuation, affecting the sensor reading; as part of the design (see Section~\ref{ssec:plcimposed}). }  $a_1-a_3$, and mapped to sensor readings via $f(a_1,\cdots,a_8$) = $s_{reading}$, provided in the normal dataset. Note, only the input actuator state space is reduced. The sensor reading remains a function of all related actuators (see Fig.~\ref{fig:truthtable}). Sensor reading is readily available from the normal dataset and its measured range, w.r.t. $a_1,\cdots,a_8$ actuation remains unchanged. In short, we refined our search space to the sensor's nn-actuator(s). However, the set of normal operations seen in the normal dataset might be a tiny subset of the entire normal state space. In this context, everything not seen as part of the normal plant operation dataset and derivations are treated as anomalous (similar to the concept used in one-class classification~\cite{Seliya2021, Cao2021}), and our detector flags it as a warning. The detector warnings provide an additional layer of safety along with the PLC boundaries discussed in section~\ref{ssec:boundary},  by generating a suitable alert when action is required.

	\begin{figure*}
		\centering
		\includegraphics[width=0.75\linewidth]{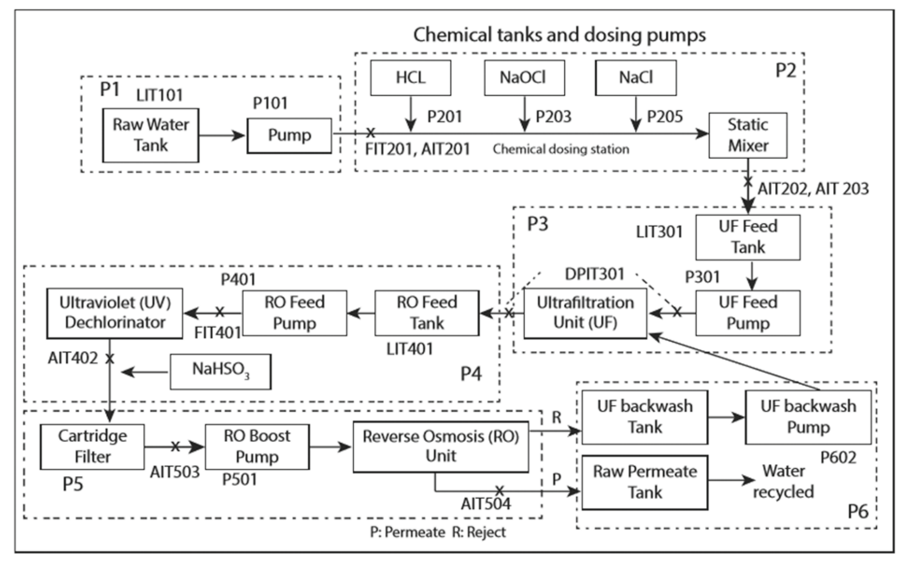}
		\caption{\small High level system architecture of SWaT water treatment testbed ~\cite{Goh2016ADT}.}
		\label{fig:swat}
	\end{figure*}

	\subsection{SWaT Water Treatment Testbed and datasets}
	\label{ssec:swat}

	The SWaT water treatment (see Fig.~\ref{fig:swat}) is  a  six-stage process ($P_1$ to $P_6$) and emulates a typical treatment plant. The water for treatment is the input to $P_1$ and treated water is the main output of $P_6$. Different (and redundant) sensors are strategically placed to make measurements across the testbed. Actuators are used for control. The sensors and actuators used in $P_1-P_6$ appear in TABLE \ref{table:device}.
	Process $P_1$ is designed to get the raw water into the system.
	$P_2$ is focused on chemical dosing and $P_3$ with ultra-filtration.
	$P_4$ includes UV-dechlorination and $P_5$ carries out reverse osmosis (RO).
	$P_6$ uses a backwash tank for cleaning and stores treated water in another tank (that is recycled). 
	The device names consist of two parts where the first 3 characters are shortened for devices type and the last 3 numbers refer to the process stage and number of devices. For example, the device FIT-101 refers to the flow meter in the first process where it is the first flow meter. 
	This description of each device can be found as follows: FIT-XXX: Flow meter, LIT-XXX: Level Transmitter, AIT-XXX: Analyser (3 types to measure ORP, pH snd conductivity), DPIT-XXX:  Differential pressure indicating transmitter, PIT-XXX: Pressure meter, MV-XXX: Motorized valve, P-XXX: Pump and UV-XXX: UV Dechlorinator~\cite{Goh2016ADT}.
	The dataset was collected for eleven days, of which the first seven days is under normal working conditions. The final four days included injecting/ swapping with anomalous data, involving 41 CPS attacks on the SWaT water testbed. The plant was run from an empty to a fully operational state. The attacks were conducted by altering the OT network traffic, spoofing the sensor values, and issuing bogus SCADA commands.

	\begin{table}[ht]
		\centering
		\resizebox{\columnwidth}{!}{
			\begin{tabular}{ccc}
				\hline
				\textbf{\textbf{Process}} & \textbf{\textbf{Sensor}} & \textbf{\textbf{Actuator}} \\ \hline
				\textbf{P1} & LIT-101, FIT-101 & MV-101, P101 \\ \hline
				\textbf{P2} & \begin{tabular}[c]{@{}c@{}}AIT-201, AIT-202, \\ AIT-203, FIT-201\end{tabular} & \begin{tabular}[c]{@{}c@{}}MV-201, P-201, P-202, P-203, \\ P-204, P-205, P-206\end{tabular} \\ \hline
				\textbf{P3} & DPIT-301, FIT-301, LIT-301 & \begin{tabular}[c]{@{}c@{}}MV-301, MV-302, MV-303, MV-304, \\ P-301, P-302\end{tabular} \\ \hline
				\textbf{P4} & \begin{tabular}[c]{@{}c@{}}AIT-401, AIT-402, \\ FIT-401, LIT-401\end{tabular} & \begin{tabular}[c]{@{}c@{}}P-401, P-402, P-403, \\ P-404, UV-401\end{tabular} \\ \hline
				\textbf{P5} & \begin{tabular}[c]{@{}c@{}}AIT-501, AIT-502, AIT-503, AIT-504,\\ FIT-501, FIT-502, FIT-503, FIT-504, \\ PIT-501, PIT-502, PIT-503\end{tabular} & P-501, P-502 \\ \hline
				\textbf{P6} & FIT-601 & P-601, P-602, P-603 \\ \hline
			\end{tabular}
		}
		\vspace{0.05cm}
		\caption{\small The sensors and actuators in each process stage\,\cite{Mathuros2024}.}
		\label{table:device}
	\end{table}
	
	The SWaT datasets provided the state view of the fifty-one sensors and actuators in the plant at one second resolution, as time series data.
	A state may include actuators in either open/close, on/off or open/transition/close state, depending on the type of the actuator used. 
	The normal dataset consisted of 495,000 records of plant state under normal operations.
	The attack dataset had 449,919 records. Note that the attack dataset is a mix of normal and attack records. The attack records accounted for were less than 6\% of total data~\cite{Mathuros2024}. If the plant was under attack during the record generation (within the dataset) then this record is tagged as \textit{Attack}, and otherwise as \textit{Normal}.
	
	\subsubsection{Sensor boundary independence in SWaT}
	\label{sssec:sensorindependence}
	SWaT has seven types of sensors --- to measure tank level, flow rate, ORP, conductivity, pH, pressure and differential pressure. 
	An anomaly detector may be designed to individually test the operational boundaries on the sensor within a process; without attempting to correlate its readings with other sensors in the same or different processes. Therefore, though the processes are interdependent, the individual sensor safety boundaries may be determined independently from its normal operation readings. 
	When the sensor value is within determined boundaries, it is deemed safe w.r.t. the metric it measures, within the SWaT process. 
	Hence, it is sufficient to look at the local causal paths to determine the sensor boundaries.
	When a sensor reading was spoofed to read within normal bounds but was actually out of bounds, we rely on other individual related sensors (if any)  to report surpassing  its boundary threshold.
	This notion of per sensor boundary computation makes the proposed anomaly detector scalable to large (mostly) serial systems. Any sensor noise will be considered as part of normal sensor behavior (provided they are within OEM tolerance limits). The data read out by the sensor itself may be linear or nonlinear. We are only interested to capture its normal operations bounds.   ICS behavior may vary over long production cycles (spanning months or years) with maintenance windows and sensor drift. These are out of scope of this work.

	\subsection{Threat Model}
	\label{ssec:threatmodel}
	An attacker is assumed to have followed an attack vector~\cite{CAPEC2025} to gain access into plant OT. This adversary is assumed to be able to inject control commands and modify sensor readings and actuator status in OT.
	This leads to four attack types~\cite{Goh2016ADT} --- single stage single point (SSSP) where exactly one sensor/actuator data is under attack, single stage multi-point (SSMP) where multiple sensor/actuator data in exactly one process is under attack, multistage single point (MSSP) where exactly one sensor/actuator data in multiple processes
	are under attack, and multistage multi-point (MSMP) where multiple sensor/actuator data in multiple processes are under attack.
	It is assumed there are no attacks on the plant when the normal dataset is collected, which would be used as the training dataset.
	
	\section{Giant-Step Baby-step Solution}
	\label{sec:solution}

	\begin{figure}
		\centering
		\includegraphics[width=0.95\linewidth]{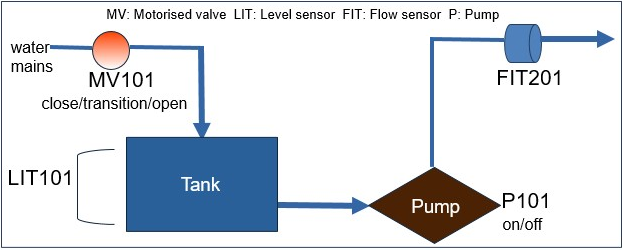}
		\caption{\small Mid-granular level view of $P_1$ and early $P_2$ SWaT process stages.}
		\label{fig:swat_stage1}
	\end{figure}

	\subsection{Design rationale and considerations}
	\label{ssec:design}
	
    The sensor and actuator values in the training dataset are viewed as ground truth data.
	Our core requirements~\cite{RiegerUsenix203} are (R1) near real-time detection, (R2) anomaly cause identification, and (R3) minimising false alarms. 
	An automated response is not a core consideration because governance considerations involve human intervention and human-in-the-loop solutions~\cite{Bentley2024,CLEMMENSEN2025112348} to be qualified as compliant to safety standards i.e. a fully automated solution might not be viable. While the ability to automate is provided, the level of automation is left to the discretion of the implementer.
	One-class classification is used due to the typical imbalance between normal and attack dataset~\cite{Yuan2023}; the latter contributing only a tiny percentage of attack data. We train using the normal dataset and employ unsupervised learning~\cite{Mujeeb2020}. 
	
	\subsection{Solution outline}
	\label{ssec:outline}
	To meet the requirements in Section~\ref{ssec:design}, we use binary classification to solve a decision problem~\cite{Chakrabarty2019}.
	The proposed anomaly detector has two main steps --- giant and baby step. The giant-step is used to determine [LB, UB] boundary for measured sensor values (available in normal dataset), in relation with every nn-actuation state. 
	For example, consider the more detailed SWaT process stages as illustrated in Fig.~\ref{fig:swat_stage1}. In this system, water flows into the tank when motorized valve (actuator) MV101 transitions from close to open. The tank level indicator (sensor) LIT101 measures the water tanks instantaneous water level. When pump (actuator) P101 is turned on, water is pumped out from the tank, and the flow transmission (sensor) FIT201 registers non-zero flow values. 
	The two nn-actuators for sensor LIT101 are MV101 and P101. MV101 has three actuation states (close-1/transition-0/open-2) and P101 has two actuation states (on-2/off-1), in combination this allows for six possible states across the two actuators connected to the sensor LIT101. The $[LB_i, UB_i]_{i=1,\cdots,6}$  bounds for LIT101 is computed by our giant-step solution for all actuator combinations of MV101 and P101: 11, 12, 01, 02, 21, and 22. 
	
	The baby-step computes the [LB, UB]
	boundary based upon the \textit{rate of change} in sensor reading(s) associated with its nn-actuator(s).
	Again, consider the level indicator sensor LIT101 in Fig.~\ref{fig:swat_stage1}. Instead of directly measuring the instantaneous level value of water in the tank, here we measure the rate of change of water level between consecutive time intervals. As seen earlier, we find $[LB_i, UB_i]_{i=1,\cdots,6}$ bounds for LIT101 in relation to actuator combinations of MV101 and P101. 
	These measurements form the basis of the min-max bounds for the rate of change of water level; for each of the six actuation states w.r.t. the level sensor.
	The combination of the proposed giant-step and baby-step methods is not only able to detect sensor values that are out of bound for dependent actuator states but also detect whether the rate of change of sensor value is also within bounds. 
	When the readings from the sensor under observation are beyond determined thresholds/bounds, the detector issues a warning event to a suitable log; to be processed and considered as a trigger for the generation of an alert to system operators.

	\begin{figure}
		\centering
		\includegraphics[width=0.95\linewidth]{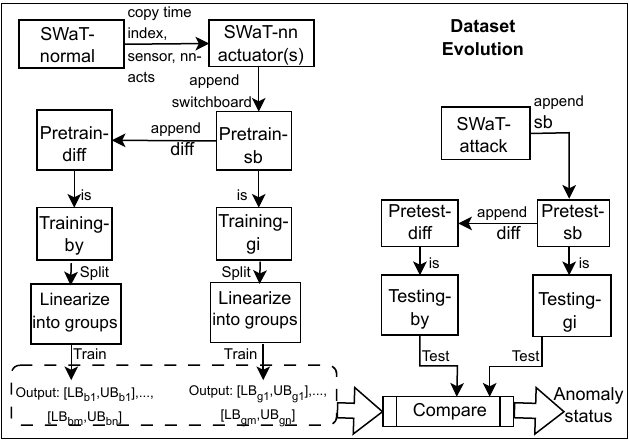}
		\caption{\small Evolution of training and testing dataset. Pretest, testing  and comparison are carried out in real-time, once the output from training is available.}
		\label{fig:evolution}
	\end{figure}

	\subsection{A switchboard to map actuation state, building linearized state groups, and enable explanation}
	\label{ssec:switchboard}
	
	The training dataset is built on the normal SWaT dataset and the testing dataset from the SWaT attack dataset (see Fig.~\ref{fig:evolution}). 
	Dataset prefixed with \textit{example} are hypothetical and are used to drive the core idea (see TABLE~\ref{table: switch_tab} and \ref{table: attackrow}).
	
	\subsubsection{Switchboard}
	A switchboard is used to linearize actuation states. It is also part of a bijective map between the data output from training  and the testing (sensor) value. 
	A switchboard state ($sb$) is a numerical string. It uniquely identifies the nn-actuator(s) state for the sensor, at time index $t$, and is expressed as a concatenation of the nn-actuator state. 
	Consider two nn-actuators, one with 3 possible states (off = 1, transition = 0, on = 2), and the other with 2 possible states (off = 1, on = 2).
	This results in six switchboard states --- 11, 12, 01, 02, 21, and 22.
	First, we parse through the normal training dataset and copy the time index, sensor reading, nn-actuators (see first four columns in example TABLE~\ref{table: switch_tab}).
	Next, we append a switchboard state (for nn-actuator(s)), at each time index $t$. This is called the pretrain-sb dataset (see additional column labeled \texttt{Switchboard} in TABLE~\ref{table: switch_tab}).
	Next, for the baby-step method, we compute the difference (diff) between consecutive sensor values and append it to a new column. We call this \textit{Pretrain-diff} dataset (see additional column labeled \texttt{LIT101Diff} in TABLE~\ref{table: switch_tab}). This is also the \textit{Training-by} dataset.
	For the giant-step method, the direct sensor values are used. Hence, \textit{Training-gi} and \textit{Pretrain-sb} datasets are identical (see Fig.~\ref{fig:evolution}).

	\begin{figure}
		\centering
		\includegraphics[width=0.95\linewidth]{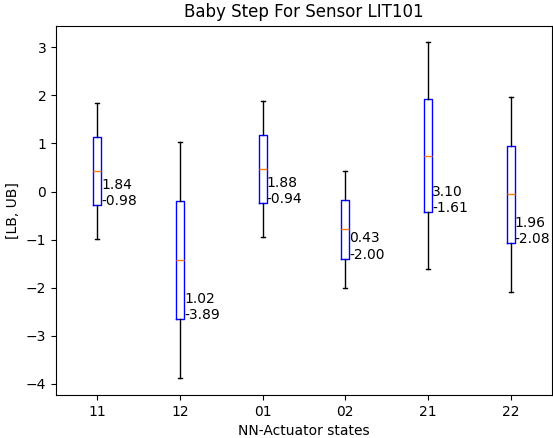}
		\caption{\small Plot of six [LB, UB] baby-step training bounds for sensor LIT101.  SWaT normal dataset is used for training.}
		\label{fig:LIT101Baby}
	\end{figure}
	
	\subsubsection{Linearized state groups} 
	Further, we split the \textit{Training-by} dataset into linearized groups ($LSG_{sb}$), for all $sb\in$ \textit{Training-by} dataset. 
	The first group contains only dataset rows associated with the sb = 11 switchboard state for this example, in the exact order it appeared in the \textit{Training-by} dataset. The remaining groups are determined in a similar manner. 
	We are now able to study the behaviour within each linearized group.
	The output of passing the training dataset through 1) Switchboard, and 2) Linearized state groups, are a set of [LB, UB] state bounds, to be used in anomaly detection (see Fig.~\ref{fig:evolution}). 
	A similar linearization split is carried out for \textit{Training-gi} dataset, and another set of [LB, UB] state bounds are determined. This completes the training steps. 
	For testing, the test sensor value $testval$ (with corresponding $sb$ state appended to it) is read from the testing dataset.

	\begin{algorithm}[t]
		\algsetup{linenosize=\tiny}
		\caption{Giant-step Baby-step Training}
		\label{alg:giby}
		\DontPrintSemicolon
		
		\footnotesize
		\SetKwProg{Fn}{Function}{:}{\KwRet}
		\SetKwFunction{NEARESTNEIGHBORS}{NearestNeighbors}
		\SetKwFunction{DATASETDIFF}{DatasetDiff}
		\SetKwFunction{SWITCHBOARDSTATE}{SwitchBoardState}
		\SetKwFunction{LINEARIZESTATES}{LinearizeStates}
		\SetKwFunction{DETERMINEBOUNDS}{DetermineBounds}
		\SetKwFunction{GIANTSTEPTRAIN}{GiantStepTrain}
		\SetKwFunction{BABYSTEPTRAIN}{BabyStepTrain}

		\KwInput{Set1: Normal SWaT dataset (NSD). Set2: Sensor label $s$ and time index $i$. Set3: Sensors-actuators relationship graph $RG$.}
		\KwOutput{The state bounds [LB, UB] for the sensor.}
		
		\SetKwProg{Fn}{Def}{:}{}
		\Fn{\NEARESTNEIGHBORS{$s, RG$}}{	
			return nearestActsList:= FindInOutSensorEdges($s$)
		}

		\Fn{\SWITCHBOARDSTATE{$i,s, actstate,RG$}}{	
			nnActs:= NearestNeighbors($s,RG$)\\
			sbFlag:= isMember($actstate, nnActs$) \\
			if sbFlag==True: return sb:= Concat(nnActs)\\
			else: return -1 
		}

		\Fn{\DATASETDIFF{$NSD,s$}}{
			\For{index $i=1$ to  end {\tcp{$i\in NSD$}}}{
				sensorval:= FindSensorVal($NSD,i,s$)\\
				diffSenVal:= sensorval[i]-sensorval[i-1]\\
				NSD[i]:= AppendToRow($NSD[i],diffSenVal$)\\
			}
			
			return NSD
		}
		
		\Fn{\LINEARIZESTATES{$s,NSD, RG$}}{	
			\For{index $i=0$ to  end {\tcp{$i\in NSD$}}}{
				nnActs:= NearestNeighbors($s,RG$)\\
				senval:= FindSensorVal($NSD,i,s$)\\
				actstate:= FindAllActState($i,s,NSD$)\\
				sb:=SwitchBoardState($i,s,actstate,RG$)\\
				LSG[sb]:= AppendRow($i,s, senval, nnActs, sb$)\\
			}
			return LSG\tcp{Linearized State Group}
		}

		\Fn{\DETERMINEBOUNDS{$LSG$}}{	
			\For{Each linearized state group LSG}{
				LB:= min($senval_i$)$\forall$ $i\in$ LSG\\
				UB:= max($senval_i$)$\forall$ $i\in$ LSG\\
				LSGBoundList:= AppendRow($s,sb,LB,UB$)\\
			}
			return LSGBoundList
		}

		\Fn{\GIANTSTEPTRAIN{$s,NSD, RG$}}{	
			LSG:=LinearizeStates($s,NSD, RG$)\\
			GiBoundList:= DetermineBounds($LSG$)\\
			
			return GiBoundList
		}

		\Fn{\BABYSTEPTRAIN{$s,NSD, RG$}}{
			NSDdiff:= DatasetDiff(NSD,s)\\	
			LSG:=LinearizeStates($s,NSDdiff, RG$)\\
			ByBoundList:= DetermineBounds($LSG$)\\
			
			return ByBoundList
		}	
	\end{algorithm}

	\begin{algorithm}[t]
		\caption{Giant-step Baby-step Testing}
		\label{alg:testing}
		\DontPrintSemicolon
		
		\footnotesize
		\SetKwProg{Fn}{Function}{:}{\KwRet}
		\SetKwFunction{BOUNDSCHECK}{BoundsCheck}
		\SetKwFunction{GIANTSTEPTEST}{GiantStepTest}
		\SetKwFunction{BABYSTEPTEST}{BabyStepTest}

		\KwInput{Set1: Attack SWaT dataset (ASD). Set2: Sensor label $s$ and time index $i$. Set3: Sensors-actuators relationship graph $RG$ and training bounds lists.}
		\KwOutput{The state bounds [LB, UB] for the sensor.}
		
		\SetKwProg{Fn}{Def}{:}{}
		
		\Fn{\BOUNDSCHECK{$i,s,senval,actstate,RG,BoundList$}}{
			sb:=SwitchBoardState($i,s,actstate,RG$)\\
			Bounds:= retrieveTrainingBounds($s,sb, BoundList$)\\
			LB:=Bounds['Low']\\
			UB:=Bounds['High']\\
			\If{LB$<=$senval$<=$UB ==False or sb==-1}{
				print(“Explanation: Anomaly DETECTED for sensor $\underline{s}$ at time index $\underline{i}$ for actuation state $\underline{actstate}$ because sensor value $\underline{senval}$ not in $[\underline{LB},\underline{UB}]$. Otherwise, actuation state is invalid or not seen in training when \underline{$sb$} = -1.")}
			\Else{ 
				print(``No anomaly was detected.")}
		}

		\Fn{\GIANTSTEPTEST{$i,s,ASD, RG, GiBoundList$}}{
			actstate:= FindAllActState($i,s,ASD$)\\
			senval:= FindSensorVal($ASD,i,s$)\\
			BoundsCheck($i,s, senval, actstate,RG, GiBoundList$)
		}

		\Fn{\BABYSTEPTEST{$i,s,ASD, RG,ByBoundList$}}{
			actstate:= FindAllActState($i,s,ASD$)\\
			senvalNow:= FindSensorVal($ASD,i,s$)\\
			senvalPrev:= FindSensorVal($ASD,i-1,s$)\\
			senval: = senvalNow - senvalPrev\\
			BoundsCheck($i,s, senval, actstate, RG, ByBoundList$)
		}	
	\end{algorithm}

	\subsubsection{Explainability}
	The switchboard provides bijective mapping from training dataset to testing dataset, using a unique numerical string  to represent nn-actuation state. The [LB, UB] bounds are determined from the linearized groups. The
	bijective nature implies the existence of an inverse mapping, from the testing set to training set (see Fig.~\ref{fig:evolution}). I.e., provided with the nn-actuator(s) and the test sensor value, we can determine its corresponding training [LB, UB] bounds for the giant and baby-step. When the sensor value is out of bounds, the explainability includes sensor label/name, actuation state, the time index \textit{t}, and the bound it breached (LB or UB). Pseudo-code for the explainability method is included within Algorithm~\ref{alg:testing}.

	\subsection{An example of operations}
	\label{ssec:exampleofoperations}
	
	In TABLE~\ref{table: switch_tab}, the sensor considered is LIT101. Its nn-actuators are MV101 and P101 (see Fig.~\ref{fig:swat_stage1}). The switchboard state at  $t$  is an ordered concatenation of its nn-actuator state. LIT101Diff is the difference between consecutive LIT101 readings. Assume first four columns in TABLE~\ref{table: switch_tab} were the only entries in the normal dataset. 
	The remaining two columns are derived using information from the earlier columns. 
	As part of the training, the linearized groups $LSG_{sb}$ are computed for each of existing switchboard states on the training dataset.
	Let $btr$ denote the baby-step training. 
	The $btr$ step computes the min and max boundaries,  $[LB,UB]_{sb=11}^{btr}$ = [0.0011, 0.1570] and $[LB,UB]_{sb=01}^{btr}$ = [0.0785, 0.4711] (see TABLE~\ref{table: switch_tab}). 
	These trained $[LB, UB]$ bounds are used to determine the anomaly status. 
	\begin{table*}
		\centering
		\begin{tabular}{cccccc}
			\hline
			\textbf{Index} & \textbf{LIT101} & \textbf{MV101} & \textbf{P101} & \textbf{Switchboard} & \textbf{LIT101Diff} \\ \hline
			1&	121.2518&	1&	1& 11  & \_ \\ \hline
			2&	121.4088&	1&	1& 11  & 0.1570 \\ \hline
			3&	121.4099&	1&	1& 11  & 0.0011 \\ \hline
			4&	121.6050&	0&	1& 01  & 0.1951 \\ \hline
			5&	121.6835&	0&	1& 01  & 0.0785 \\ \hline
			6&	122.1546&	0&	1& 01  & 0.4711 \\ \hline
		\end{tabular}
		\vspace{0.05cm}
		\caption{\small An example normal (training) dataset for sensor LIT101 and its nn-actuators. Switchboard and diff values are derived.}
		\label{table: switch_tab}
	\end{table*}
	\begin{table*}
		\centering
		\begin{tabular}{cccccc}
			\hline
			\textbf{Index} & \textbf{LIT101} & \textbf{MV101} & \textbf{P101} & \textbf{Switchboard} & \textbf{LIT101Diff} \\ \hline
			1&	123.2151&	1&	1& 11  & \_ \\ \hline
			2&	121.6835&	1&	1& 11  & -1.5316 \\ \hline
		\end{tabular}
		\vspace{0.05cm}
		\caption{\small An example attack (testing) dataset showing row entries for sensor LIT101 and its nn-actuators. }
		\label{table: attackrow}
	\end{table*}
	
	An example attack (testing) dataset (see first four columns) is shown in TABLE~\ref{table: attackrow}. 
	As seen earlier, the remaining  two columns are derived.
	The testing switchboard state $sb=11$ is mapped to matching  $[LB,UB]_{sb}^{btr}$, seen in training.
	The corresponding testing sensor value is checked to determine if it satisfies the condition $LB\leq testsensorval\leq UB$.
	For time index 1 in TABLE~\ref{table: attackrow}, LIT101Diff is undefined. This is because computing a diff, i.e., current value - previous value, requires two consecutive (time indexed) LIT101 sensor values. For testing index 2, a diff exists. With respect to corresponding (training) $sb=11$ actuation state, its baby-step bounds are [0.0011, 0.1570].
	The verification check therefore is 0.0011$\leq$LIT10Diff$\leq$0.1570?   Which in our example (LIT10Diff = -1.5316) returns the \textit{False} condition, resulting in the baby-step method issuing an out of bounds warning event.
	The 6 baby-step bounds for LIT101 determined from training using the SWaT normal dataset are shown in Fig.~\ref{fig:LIT101Baby}.
	Had the giant-step been used, we would train and test directly on the sensor values in column labeled \texttt{LIT101} instead of column labeled \texttt{LIT101Diff}, in TABLE~\ref{table: switch_tab} and TABLE~\ref{table: attackrow}, respectively. 
	
	For other sensors such as FIT201 (see Fig.~\ref{fig:swat_stage1}), it is of interest to find the giant and baby-step training bounds. When water is pumped using P101, the flow rate measured by FIT201 settles in a steady range of (mostly) adjacent values. Training directly on the flow rate, and finding bounds, allows us to detect anomalous flows. The baby-step is also of significance w.r.t. this sensor. We are also interested in the bounds on rate of flow change, w.r.t. its nn-neighbor(s) actuation.
	Later, in section~\ref{sec:extended} we will see that the baby and giant-step may be combined with an extended detection algorithm, to provide up to four different detection mechanisms.

	\begin{figure*}
		\centering
		\includegraphics[width=1.0\linewidth]{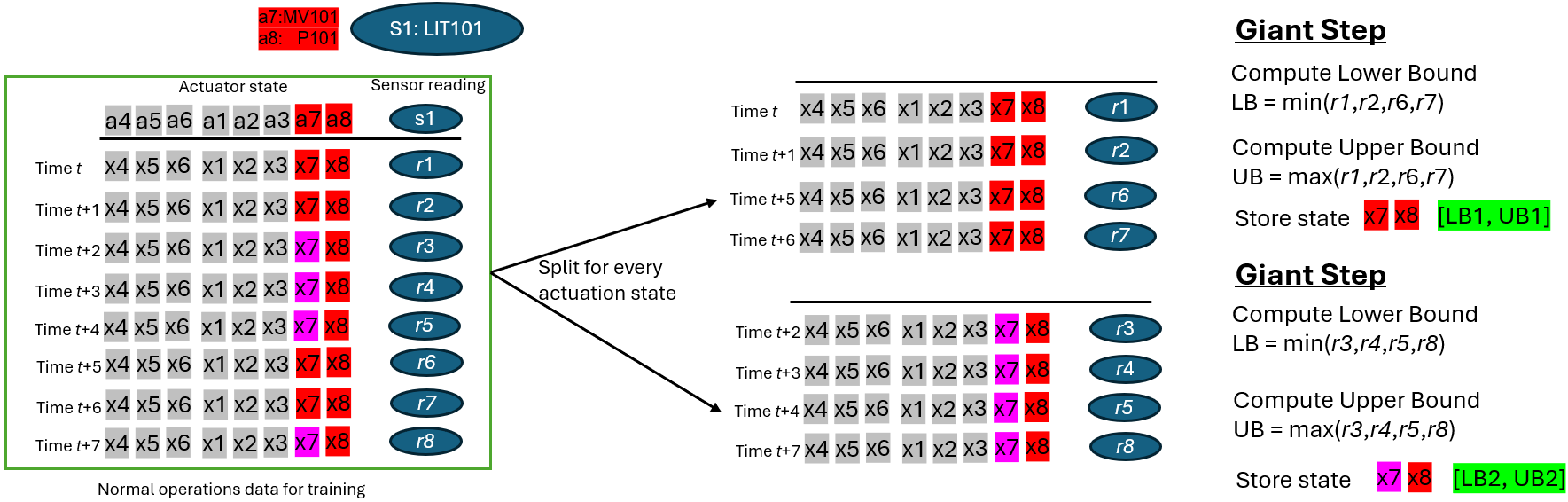}
		\caption{\small A SWaT process $P_1$ water pumping example.
			In Fig.~\ref{fig:swat_stage1}, the sensor considered is the level indicator sensor LIT101. Its nearest neighbor actuators are a7 = MV101 and a8 = P101. Actuation states for a1-a6 are absent in this example. At times $t+2$, $t+5$, and $t+7$, the actuation state of a7 has changed. For brevity  only binary actuation is shown. In practice, the motorized valve MV101 transitions from OPEN to CLOSE and vice versa in a given number of time steps; resulting in corresponding transitional sensor readings. Here, only the steady-state readings are shown.
			For our experiments, transition states are treated as separate states.
			The Giant-step training splits the time-series data for every actuation state seen. It then computes and stores the minimum and maximum values seen for the actuation state; for the given sensor. }
		\label{fig:giantstepgraphics}
	\end{figure*}
	
	\begin{figure*}
		\centering
		\includegraphics[width=1.0\linewidth]{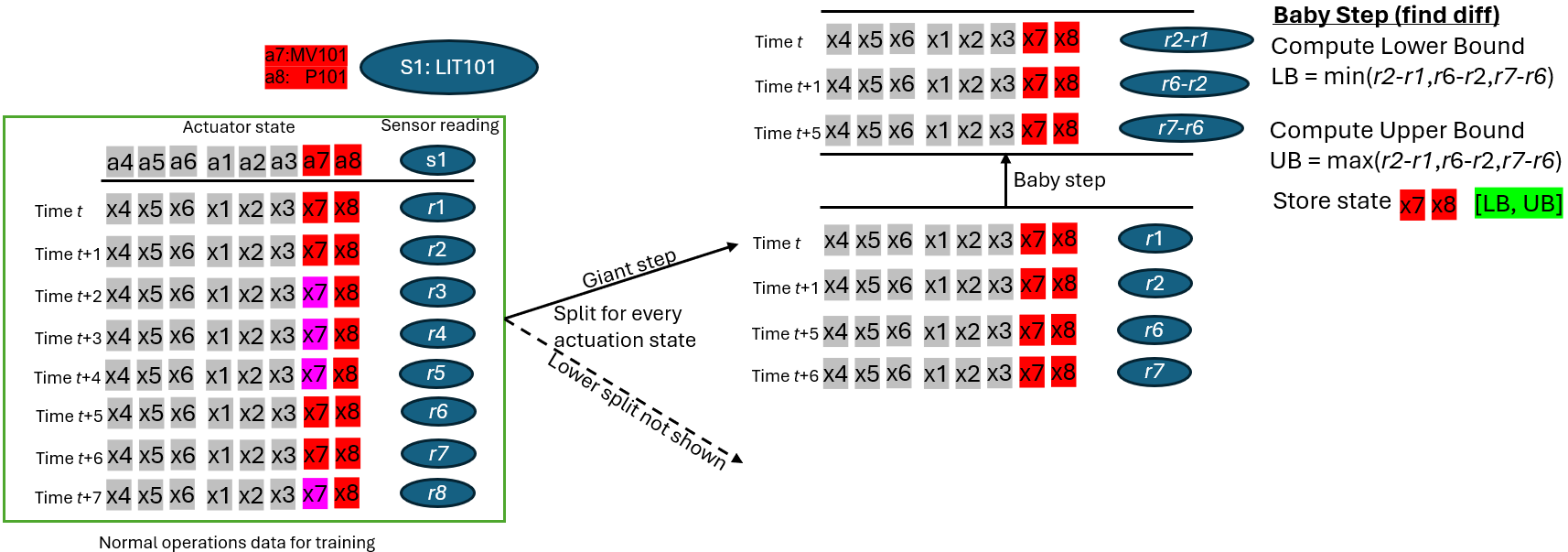}
		\caption{\small The  Baby-step training split is similar to the Giant-step shown in Fig.~\ref{fig:giantstepgraphics}. However, after the split, it computes the difference between adjacent current and previous sensor readings.
			It then proceeds to compute the bounds and stores it as part of training output.		
		}
		\label{fig:babystepgraphics}
	\end{figure*}
	
	\begin{figure*}
		\centering
		\includegraphics[width=1.0\linewidth]{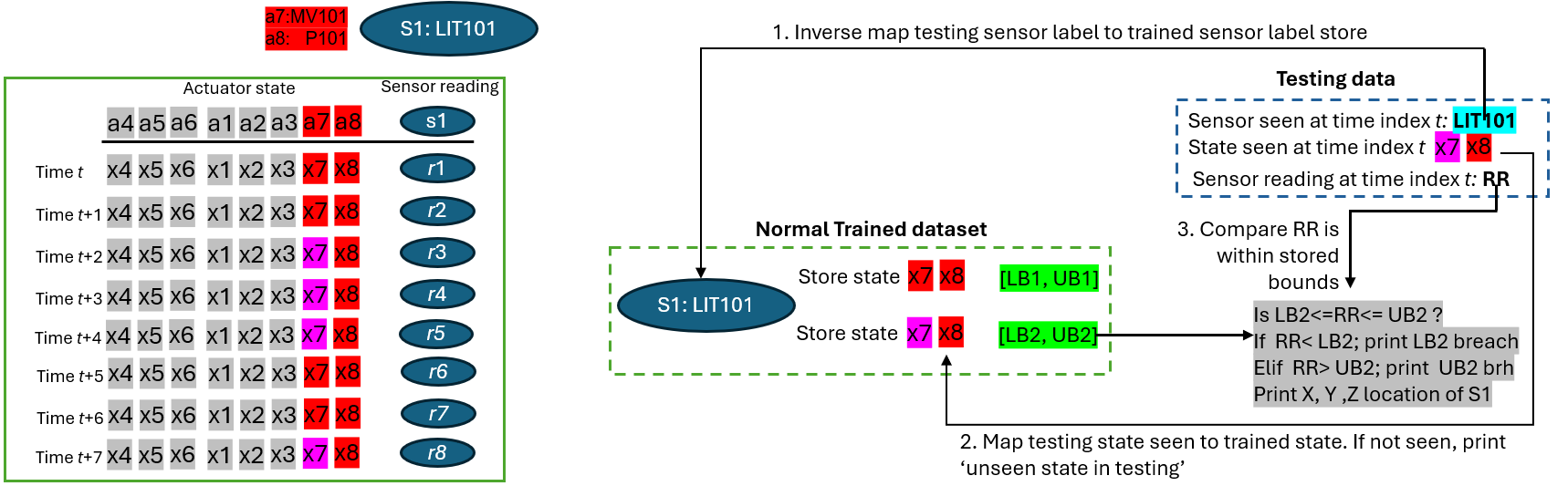}
		\caption{\small This example shows testing and explainability for the sensor trained in Giant-step (Fig.~\ref{fig:giantstepgraphics}) and Baby-step (Fig.~\ref{fig:babystepgraphics}). For testing w.r.t. Giant-step, the direct sensor reading is considered. For the baby-step testing, the difference between adjacent readings is taken. 1. The first step is to inverse map the testing sensor label to the training sensor label. 2. Next, we map the testing actuation state seen to the training actuation state of the sensor considered. The training sensor label store holds the bounds for all seen actuation states. Once the testing actuation state is mapped to the training actuation state; the bounds are read out from the store. 3. Next, the testing sensor value at time $t$ is verified against the bounds and an appropriate message is raised --- depending on whether it is within or out of bounds. Any out of bound result further triggers a read out of the location of the sensor in X, Y, Z coordinates in the floor plan of the plant. The inverse mapping makes testing extremely fast and instantaneously points the operator to the anomalous sensor location.     }
		\label{fig:testandexplaingraphics}
	\end{figure*}
	
	\subsection{Giant-Step Baby-Step Anomaly Detector }
	\label{ssec:giby}
	The  \textbf{Gi}ant-step \textbf{B}ab\textbf{y}-step (GiBy) anomaly detector is based on the switchboard implementation, linearization step, and determination of training bounds detailed in Section~\ref{ssec:switchboard}. 
	The algorithm is run for each sensor with a nn-actuator(s) dependency. 
	We note that the GiBy algorithm is able to take as input, nn-actuator(s) with different actuation sequences. For example, the first actuator $a_1$ might have $|k_1|=2$, $a_2$ with  $|k_2|=3$ and $a_3$ actuator with $|k_3|=2$.
	There is also no limitation on the number of input nn-actuators imposed by the algorithm.  
	Both, Algorithm~\ref{alg:giby} and Algorithm~\ref{alg:testing} shows GiBy-core training and testing, respectively, for one sensor.
	Using an example, the giant-step training is  visually explained in Fig.~\ref{fig:giantstepgraphics}, the baby-step training in Fig.~\ref{fig:babystepgraphics}, and the testing and explainability in Fig.~\ref{fig:testandexplaingraphics}.

	\begin{table*}[ht]
	\resizebox{\linewidth}{!}{%
		\begin{tabular}{cccccc}
			\hline
			\textbf{Attack no.} & \textbf{Attack point} & \textbf{Attack event} & \textbf{Detected} & \textbf{nn-actuator(s); sensor} & \textbf{\begin{tabular}[c]{@{}c@{}}Remark\end{tabular}} \\ \hline
			1 & MV-101 & Open MV-101 & Y & MV-101; LIT-101 & Detected with Giant-step \\ \hline
			2 & P-102 & Turn on P-102 & Y & P-101, P-102; FIT201 & Detected because state not seen in training \\ \hline
			3 & LIT-101 & Increase by 1 mm every second & Y, N & MV-101, P-101; LIT-101 & One of two attack states detected with Baby-step extended algorithm \\ \hline
			4 & MV-504 & Open MV-504 & N & \_ & MV-504 readings not in SWaT datasets \\ \hline
			6 & AIT-202 & Set value of AIT-202 as 6 & Y & P-101, MV-201; AIT-202 & Detected with Giant-step\\ \hline
			7 & LIT-301 & Water level increased above High-High set point & Y & MV-201; LIT-301 & Detected with Baby-step  \\ \hline
			8 & DPIT-301 & Set value of DPIT as \textgreater{}40kpa & Y & P-301; DPIT-301 & Detected with Giant-step \\ \hline
			10 & FIT-401 & Set value of FIT-401 as \textless{}0.7 & Y & P-401, UV-401; FIT-401 & Detected with Baby-step \\ \hline
			11 & FIT-401 & Set value of  FIT-401 as 0 & Y & P-401, UV-401; FIT-401 & Detected with Baby-step \\ \hline
			13 & MV-304 & Close MV-304 & N & MV-302, MV-304; DPIT-301 & DPIT-301 reads normal. No other  sensor relates directly with state \\ \hline
			14 & MV-303 & Do not let  MV-303 open & N & MV-303, MV-304; DPIT-301 & DPIT-301 reads normal. No other sensor relates directly with state  \\ \hline
			16 & LIT-301 & \begin{tabular}[c]{@{}c@{}}Decrease water level \\ by 1mm each second\end{tabular} & Y & MV-201; LIT-301 & Detected with Baby-step extended algorithm \\ \hline
			17 & MV-303 & Do not let MV-303 open & N & MV-303, MV-304; DPIT-301 & DPIT-301 reads normal. No other  sensor relates directly with state \\ \hline
			19 & AIT-504 & Set value of  AIT-504 to 16 $\mu$S/cm & Y & P-501, P-502; AIT-504 & Detected with Giant-step  \\ \hline
			20 & AIT-504 & Set value of AIT-504 to 255 $\mu$S/cm & Y & P-501, P-502; AIT-504 & Detected with Giant-step\\ \hline
			21 & MV-101, LIT-101 & \begin{tabular}[c]{@{}c@{}}Keep MV-101 on continuously; \\ Value of LIT-101 set as 700   mm\end{tabular} & Y & MV-101, P-101; LIT-101 & Detected with Baby-step \\ \hline
			22 & \begin{tabular}[c]{@{}c@{}}UV-401, \\ AIT-502,\\  P-501\end{tabular} & \begin{tabular}[c]{@{}c@{}}Stop UV-401, Value of AIT502\\ set as 150, Force P-501 to remain ON\end{tabular} & Y & UV-401, P-501; AIT-502 & Detected with Baby-step \\ \hline
			23 & \begin{tabular}[c]{@{}c@{}}P-602,\\ DIT-301,\\ MV-302\end{tabular} & \begin{tabular}[c]{@{}c@{}}Value of DPIT-301 set to \textgreater{}0.4 bar,\\ Keep MV-302 open, \\ Keep P-602 closed\end{tabular} & Y & MV-302, P-602; DPIT-301 & Detected with Baby-step \\ \hline
			24 & \begin{tabular}[c]{@{}c@{}}P-203,\\ P-205\end{tabular} & Turn off  P-203 and P-205 & N, Y & \begin{tabular}[c]{@{}c@{}}P-203, P-205; AIT-202\\ P-203, P-205; AIT-203\end{tabular} & \begin{tabular}[c]{@{}c@{}}Attack not detected as pH sensor values in normal range,\\ Attack detected when ORP sensor values were out of trained bounds\end{tabular}  \\ \hline
			25 & \begin{tabular}[c]{@{}c@{}}LIT-401,\\ P-402\end{tabular} & \begin{tabular}[c]{@{}c@{}}Set value of LIT-401 as 1000,\\ P-402 is kept on\end{tabular} & Y & P-402; LIT-401 & Detected with Baby-step \\ \hline
			26 & \begin{tabular}[c]{@{}c@{}}P-101, \\ LIT-301\end{tabular} & \begin{tabular}[c]{@{}c@{}}P-101 is turned on continuously,\\ Set value of LIT-301 as 801 mm\end{tabular} & Y, N & \begin{tabular}[c]{@{}c@{}}P-101, P-102; LIT-101 \\ MV-201, P-301; LIT-301\end{tabular} & \begin{tabular}[c]{@{}c@{}}Attack  detected as both pumps not simultaneously ON in training\\ Not detected as LIT301 sensor value changes were in trained bounds\end{tabular} \\ \hline
			27 & \begin{tabular}[c]{@{}c@{}}P-302, \\ LIT-401\end{tabular} & \begin{tabular}[c]{@{}c@{}}Keep P-302 on continuously\\ Value of LIT401\\  set as 600 mm till 1:26:01\end{tabular} & Y & P-302, P-402; LIT-401 & Detected with Baby-step \\ \hline
			28 & P-302 & Close P-302 & N & \_ & Turning off P-302 will stop water inflow to LIT-401 is the expected outcome \\ \hline
			29 & \begin{tabular}[c]{@{}c@{}}P-201,\\ P-203,\\ P-205\end{tabular} & \begin{tabular}[c]{@{}c@{}}Turn on P-201, Turn on P-203,\\ Turn on P-205\end{tabular} & N, N, Y &  \begin{tabular}[c]{@{}c@{}}\_\\ P203, P205; AIT202\\ P203, P205; AIT203\end{tabular} & \begin{tabular}[c]{@{}c@{}c@{}}P-201 is never turned on in training. No training data available\\pH sensor AIT-202 readings were in normal range\\Detected with Giant-step since AIT-203 readings were out of trained bounds\end{tabular} \\ \hline
			30 & \begin{tabular}[c]{@{}c@{}}LIT-101,\\ P-101,\\ MV-201\end{tabular} & \begin{tabular}[c]{@{}c@{}}Turn P-101 on continuously;\\  Turn MV-101 on continuously;\\ Set value of LIT-101 as 700 mm;\\ P-102 started itself because LIT301level \\ became low\end{tabular} & Y, Y & \begin{tabular}[c]{@{}c@{}}\\ MV-101, P-101; LIT101\\ P101, P102; FIT201\end{tabular} & \begin{tabular}[c]{@{}c@{}}Detected LIT-101 attack with Baby-step,\\ Both P-102 and P-102 not simultaneously ON in training data. This state is detected\end{tabular} \\ \hline
			31 & LIT-401 & Set LIT-401 to less than L & Y & P-302, P-402; LIT-401 & Detected with Baby-step \\ \hline
			32 & LIT-301 & Set LIT-301 to above HH & Y & MV-201; LIT-301 & Detected with Baby-step \\ \hline
			33 & LIT-101 & Set LIT-101 to above H & Y & MV-101, P-101; LIT-101 & Detected with Baby-step \\ \hline
			34 & P-101 & Turn P-101 off and P-102 ON & Y & P-101, P-102; FIT-201 & P-101 = OFF and P-102 = ON is never seen in training dataset \\ \hline
			35 & \begin{tabular}[c]{@{}c@{}}P-101,\\ P-102\end{tabular} & \begin{tabular}[c]{@{}c@{}}Turn P-101 off,\\  Keep P-102 off\end{tabular} & N & \_ & P101 = OFF and P102 = OFF, and water stop flowing is expected behavior \\ \hline
			36 & LIT-101 & \begin{tabular}[c]{@{}c@{}}Set LIT-101 \\ to less than LL\end{tabular} & Y & MV-101, P-101; LIT-101 & Detected with Baby-step \\ \hline
			37 & \begin{tabular}[c]{@{}c@{}}P-501, \\ FIT-502\end{tabular} & \begin{tabular}[c]{@{}c@{}}Close P-501, Set value of FIT-502\\ to 1.29 at 11:18:36\end{tabular} & Y & P-501; FIT-502 & Detected with Baby-step \\ \hline
			38 & \begin{tabular}[c]{@{}c@{}}AIT-402,\\ AIT-502\end{tabular} & \begin{tabular}[c]{@{}c@{}}Set value of  AIT-402 as 260,\\  Set value of  AIT-502 to 260\end{tabular} & Y, Y & \begin{tabular}[c]{@{}c@{}}UV-402; AIT-402\\ P-501, P-502; AIT-502\end{tabular} &  \begin{tabular}[c]{@{}c@{}}Attack on AIT-402 detected with Baby-step\\ Attack on AIT-502 detected with Baby-step\end{tabular} \\ \hline
			39 & \begin{tabular}[c]{@{}c@{}}FIT-401, \\ AIT-502\end{tabular} & \begin{tabular}[c]{@{}c@{}}Set value of  FIT-401 as 0.5,\\ Set value of  AIT-502 as 140 mV\end{tabular} & Y, Y & \begin{tabular}[c]{@{}c@{}}UV-401; FIT-401, \\ P-501, P-502; AIT-502\end{tabular} & \begin{tabular}[c]{@{}c@{}}Attack on FIT-401 detected with Baby-step \\ Attack on AIT-502 detected with Baby-step\end{tabular} \\ \hline
			40 & FIT-401 & Set value of  FIT-401 as 0 & Y & UV-401; FIT-401 & Detected with Baby-step \\ \hline
			41 & LIT-301 & \begin{tabular}[c]{@{}c@{}}Decrease value \\ by 0.5 mm per second\end{tabular} & Y & P-301, P-302; LIT-301 & Detected by Baby-step extended algorithm \\ \hline
		\end{tabular}%
	}\\
	\caption{\small The SWaT attacks detected using GiBy-core (Section~\ref{sec:solution}),  with extended detection capabilities; as presented in  Section~\ref{sec:extended}.}
	\label{table:att}
\end{table*}

	\section{Extending Detection Capabilities}
	\label{sec:extended}
	
\begin{figure}
		\centering
		\includegraphics[width=0.9\linewidth]{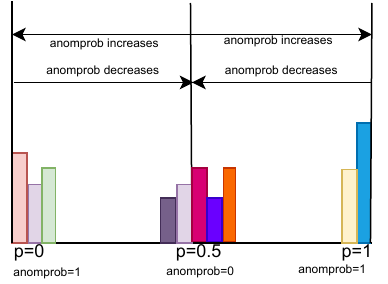}
		\caption{ \small A hypothetical illustration of the distribution probabilities for sensor readings given its nn-actuation state, centred at $p=0.5$. The anomaly probability is 0 at the centre and tends towards 1, as it approaches either of the boundaries. The x-axis is range of probability distribution and y-axis, its frequency.}
		\label{fig:extended_explanation}
	\end{figure}

	Extended capabilities are built to provide anomaly detection for a time-series window of sensor readings tending towards the boundaries or around the centre of the probability distribution (see Fig.~\ref{fig:extended_explanation}).
	Stealth attacks on ICS were studied in Urbina et al~\cite{Urbina2016}. The goal of an adversary is to keep detection statistic below
	the selected threshold. Some detection solutions using a residual-error threshold~\cite{Feng2019} are vulnerable to stealth attacks. We used stateful detection to catch anomalies based on the probability distribution of state values in a time-window. The extended detector is able to flag a subset of time-series sensor readings clustered around (within) the boundaries or towards the centre of the distribution (see Fig.~\ref{fig:extended_explanation}); those are rare events not captured during training. Note that it may be used with giant or baby-step. While it improves stealth attack detection (see Section~\ref{sec:experiments}), this extended detector may not flag attack record readouts exceptionally close to normal operations; typically, this would require an (insider) attacker with extensive prior knowledge of the system inner workings.
	
	The anomaly probability is deduced using a proposed empirical method. 
	The empirical method is preferred when a relatively large number of entries are available in the training dataset (this is case for the SWaT used within Section~\ref{ssec:swat}). The examples in TABLE~\ref{table: switch_tab1} and TABLE~\ref{table: switch_swtest} are hypothetical, and drives the core idea.
	The extended detector relies on a probability score deduced from the sensor reading. 

	\begin{figure}
		\centering
		\includegraphics[width=0.9\linewidth]{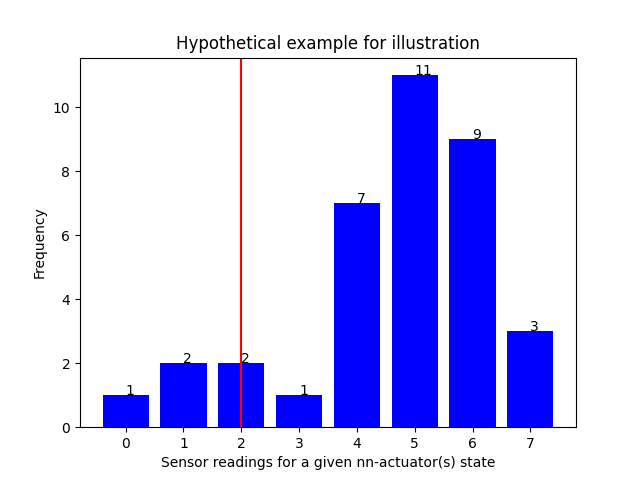}
		\caption{\small An example sensor readings distribution for a given nn-actuation state. Some readings may repeat and is represented by its y-axis frequency.}
		\label{fig:empirical_frequencies}
	\end{figure}
	
	\subsubsection{Empirical anomaly score}
	\label{sssec:empricalsssec}
	Consider the hypothetical example in Fig.~\ref{fig:empirical_frequencies}.
	The range of sensor readings is shown along the x-axis, and the frequency range of these readings is shown up the y-axis.
	The distribution of sensor values for a given sensor and a specified actuation state is shown. 
	As an example, we begin to calculate the anomaly probability for the sensor reading value of 2 (shown by the red line).
	We compute four probabilities w.r.t. the sensor --- Equations~(\ref{eqn:lprob}) and (\ref{eqn:rprob}) are the left and right anomaly probabilities, respectively. The empirical anomaly probability in shown in Equation~(\ref{eqn:panomaly}), and finally its not-anomaly probability is $1-Pr_{anom}$.
	Let $sen$ be the sensor reading of interest (for this example $sen=2$).
	It is the third reading from the left. Therefore, sensor index $n=3$.
	Let  $sen_{i}^{f}$ be the frequency at the $i^{th}$ sensor index.
	Let $l$ denote left and $r$ for right. 
	Equations~(\ref{eqn:total})-(\ref{eqn:rprob}) are shown below.
	
	\begin{equation}
		\label{eqn:total}
		T = \sum_{i=1}^{end}{sen_{i}^{f}}
	\end{equation}
	
	\begin{equation}
		\label{eqn:lprob}
		P(sen_{l}^n) = \biggl(\sum_{i=1}^{n-1}{sen_{i}^{f}}\biggr)/T
	\end{equation}

	\begin{equation}
		\label{eqn:rprob}
		P(sen_{r}^n) = \biggl(\sum_{i={n+1}}^{end}{sen_{i}^{f}}\biggr)/T
	\end{equation}

	For the example with $n=3$, we have $T=36$, $P(sen_{l}^n)$ = 3/36 = 0.083 and $P(sen_{r}^n)$ = 31/36 = 0.861.
	The center of the probability distribution is $p=0.5$.
	The anomaly probability for $sen$ is computed using Equation~(\ref{eqn:panomaly}).
	The edge cases are shown in Algorithm~\ref{alg:extendedTraining}. 
	
	\subsubsection{Sliding window in training and testing}
	\label{sssec:slidingwindowsssec}
	
	The goal is to detect a series of consecutive sensor reading probabilities whose values cluster around the probability boundaries $anomprob=1$, and  $anomprob=0$ (see example Fig.~\ref{fig:extended_explanation}).
	For training, we multiply a series of consecutive not-anomaly probabilities as part of a sliding window, for sensor readings in a time-series dataset. The min and max values $[min, max]$ of the product are computed. For testing, the same sliding window multiplication is carried out on the test dataset. For the multiplied test probability $testprob$, when the condition $min \leq testprob \leq max$ is violated, an anomaly is raised and the explanation provided.  The set of consecutive test sensor values for which anomalies are raised are seen as rare events, not captured during training of normal dataset. 
	TABLE~\ref{table: switch_tab1} shows a hypothetical training dataset for sensor LIT101 and switchboard status 11.  
	Column  labeled \texttt{1 - PrAnom}, is the empirically determined not-anomaly probability.
	PrAnom is derived using Equation~(\ref{eqn:panomaly}). Column labeled \texttt{SWProduct} is the sliding window (sw) product. The sliding window length ($sw_{len}$) was chosen to be 3, for ease of illustration. We compute Equation~(\ref{eqn:swproduct}).
	
	\begin{table*}
		\centering
		\begin{tabular}{cccccc}
			\hline
			\textbf{Index} & \textbf{LIT101} & \textbf{LIT101Diff} & \textbf{sb} & \textbf{1-PrAnom} & \textbf{SWProduct} \\ \hline
			1&	186.2518&	\_    &	11& \_      &  \_ \\ \hline
			2&	186.4088&	0.1570&	11& 0.8  &  \_ \\ \hline
			3&	186.4099&	0.0011&	11& 0.2  &  \_ \\ \hline
			4&	186.6050&	0.1951&	11& 0.6  & 0.096\\ \hline
			5&	186.6835&	0.0785&	11& 0.6  & 0.072\\ \hline
			6&	187.1546&	0.4711&	11& 0.2  & 0.072\\ \hline
		\end{tabular}
		\vspace{0.05cm}
		\caption{\small An example training dataset. NN-actuators are not shown. sb is switchboard, $SW_{len}=3$. Columns labeled \texttt{1-PrAnom} and \texttt{SWProduct} are derived.}
		\label{table: switch_tab1}
	\end{table*}

	\begin{equation}
		\label{eqn:swproduct}
		SWProduct(Index) = \prod_{i={Index-sw_{len}+1}}^{Index} 1-PrAnom_i 
	\end{equation}
	
	For the example in TABLE~\ref{table: switch_tab1}, the min and max bounds for SWProduct are [0.072, 0.096] for $sb=11$. 
	
	\begin{table*}
		\centering
		\begin{tabular}{cccccc}
			\hline
			\textbf{Index} & \textbf{LIT101} & \textbf{LIT101Diff} & \textbf{sb} & \textbf{1-PrAnom} & \textbf{SWProduct} \\ \hline
			1&	111.2324&	\_    &	11& \_      &  \_ \\ \hline
			2&	111.2824&	0.05&	11& 0.4  &  \_ \\ \hline
			3&	111.3324&	0.05&	11& 0.4  &   \_ \\ \hline
			4&	111.3824&	0.05&	11& 0.4  &   0.064\\ \hline
		\end{tabular}
		\caption{\small Test for extended anomaly. testval is LIT101Diff, $SW_{Len}=3$.}
		\label{table: switch_swtest}
	\end{table*}

	\begin{table*}
			\centering
			\begin{tabular}{cccccc}
				\hline
				\textbf{} & \textbf{nn-act(s), sen} & \textbf{diff (if any) \& Linearize}  &\textbf{Core bounds} &  \textbf{Extended bounds} & \textbf{Total time}  \\ \hline
				Giant-step & 219s & 82s & 16s & 407s & 724s \\ \hline
				Baby-step & 162s & 65s & 15s &  269s & 511s  \\ \hline
			\end{tabular}%
		\caption{\small Average time taken to train one sensor in normal dataset.}
		\label{table:timetakentrain}
	\end{table*}
	
	\begin{table*}
			\centering
			\begin{tabular}{cccccc}
				\hline
				\textbf{} & \textbf{nn-act(s), sen} & \textbf{diff (if any)}  &\textbf{Total pre-test time} &  \textbf{Total test time/\#records} & \textbf{Test time/sensor}  \\ \hline
				Giant-step & 148s & \_ & 148s & 323s/449919 &  0.00071s \\ \hline
				Baby-step & 145s & 47s & 192s &  320s/449919 & 0.00071s  \\ \hline
			\end{tabular}%
		\caption{\small Average time taken to test one sensor in attack dataset.}
		\label{table:timetakentest}
	\end{table*}

	The example test dataset in  TABLE~\ref{table: switch_swtest} shows the test sensor readings and its diff. The empirical not-anomaly probability (1-PrAnom) for LIT101Diff is 0.4, since the diffs are the same.
	The fourth test index with LIT101Diff has sliding window product, SWProduct  = 0.064. The verification check therefore is $0.072\leq SWProduct \leq 0.096$?
	Which in our example (SWProduct = 0.064) returns the \textit{False}
	condition, and anomaly is flagged by the detector. 
	For the experiments in Section~\ref{sec:experiments}, the SWaT normal dataset is
	used for training through use of its sensor values across its linearized group states (see Fig.~\ref{fig:evolution}) --- for sliding window sizes 5, 10, 25, 50 and 100. 
	For example, a sliding window of size 100 implies that 100 consecutive sensor readings for the given state at one second resolution were trained, and the SWProduct min, max values were the output of that training. This was tested against the SWaT attack dataset.  
	Different window sizes are used to capture anomalies across different time ranges and provide different detection sensitivity depending on for how long the attack lasted.
	The extended detection training methods are provided in Algorithm~\ref{alg:extendedTraining}.
	The FindMinMaxProduct function at line 49 is used to find the min and max sliding window product bounds for different window lengths and each switchboard state in LSG (derived from training dataset).

	\begin{equation}
		\label{eqn:panomaly}
		Pr_{anom}(sen)=
		\begin{cases}
			(0.5-min(P(sen_{l}^n),P(sen_{r}^n)))\cdot 2, &  \\
			\text{when}\ P(sen_{l}^n)\neq 0.5,  P(sen_{r}^n)\neq 0.5.&
		\end{cases}
	\end{equation}

	\begin{algorithm}[t]
		\caption{Extended Detection Training Functions}
		\label{alg:extendedTraining}
		\DontPrintSemicolon
		
		\footnotesize
		\SetKwProg{Fn}{Function}{:}{\KwRet}
		\SetKwFunction{SORT}{SortAscendingByValueOnFrequency}
		\SetKwFunction{TOTAL}{Total}
		\SetKwFunction{FINDPRLEFT}{FindPrLeft}
		\SetKwFunction{FINDPRRIGHT}{FindPrRight}
		\SetKwFunction{FINDANOMPR}{FindAnomPr}
		\SetKwFunction{FINDSWPROD}{FindSWProduct}
		\SetKwFunction{FINDMINMAXPROD}{FindMinMaxProduct}

		\Fn{\SORT{$LSG, sb$}}{
			{   sortedSenList:= SortAscendingByValueOnFrequency(LSG, sb)\tcp{Sort the Linearized group for switchboard state on frequency of repeated sensor values as in Fig.~\ref{fig:empirical_frequencies}.}
				return sortedSenList
		}}	
		
		\Fn{\TOTAL{$sortedSenList$}}{
			{  	
				T: = $\sum_{i=1}^{end}{sortedSenList_{i}^{f}}$}
			return T
		}	
		
		\Fn{\FINDPRLEFT{$index, sortedSenList, T$}}{
			{  n:= index\\
				PrLeft: = $\biggl(\sum_{i=1}^{n-1}{sortedSenList_{i}^{f}}\biggr)/T$}
			
			return PrLeft
		}

		\Fn{\FINDPRRIGHT{$index, sortedSenList, T$}}{
			{  n:= index\\
				end:= lastIndex(sortedSenList)\\
				PrRight: = $\biggl(\sum_{i=n+1}^{end}{sortedSenList_{i}^{f}}\biggr)/T$}
			
			return PrRight
		}	
		
		\Fn{\FINDANOMPR{$PrLeft, PrRight$}}{

			\If{PrLeft ==0 or PrRight ==0}
			{PrAnom: = 1\tcp{flag anomaly}}
			\ElseIf{PrLeft$\neq$ 0.5 and PrRight$\neq$ 0.5}
			{PrAnom:= $\mid$0.5-min(PrLeft, PrRight)$\mid\cdot 2$}
			\Else{PrAnom:= 0.5}
			
			\If{PrLeft + PrRight $<$0.5}
			{PrAnom: = $\mid$PrLeft+PrRight$\mid$}

			\If{PrRight == 0 and PrLeft $\neq$1}
			{PrAnom: = PrLeft}
			\ElseIf{PrLeft == 0 and PrRight$\neq$ 1}
			{PrAnom: = PrRight}
			return index, 1-PrAnom\tcp{Not anomaly probability}
		}

		\Fn{\FINDSWPROD{$index, swlen, LSG, sb$}}{
			{ 
				sortedSenList:= SortAscendingByValueOnFrequency(LSG, sb)\\
				T:= Total(sortedSenList)\\
				PrLeft:= FindPrLeft($index, sortedSenList, T$)\\
				PrRight:= FindPrRight($index, sortedSenList, T$)\\
				
				PrAnom:= FindAnomPr($PrLeft, PrRight$)\\
				swproduct:= $\prod_{i={Index-swlen+1}}^{Index} (1-PrAnom_i)$\\
			}
			
			return swproduct
		}

		\Fn{\FINDMINMAXPROD{$index, swlen, LSG, sb$}}{
			{ 
				\tcp{swlen 5, 10, 25, 50, 100 are used}
				end:= lastIndex(LSG, sb)\\
				minswprod:= $\inf$\tcp{A large positive number}
				maxswprod:= $-\inf$\tcp{A large negative number}
				
				\For{i=1 to end}
				{	swprod:= FindSWProduct($index, swlen, LSG, sb$)\\
					minswprod:= min(minswprod, swprod ) \\
					maxswprod:= max(maxswprod, swprod )}
			}
			
			return minswprod, maxswprod\tcp{Do for each sb state}
		}	
	\end{algorithm}

	For testing the extended algorithm, a similar sequence of steps is used but with the testing dataset.  The sensor values used here are the attack dataset readings for giant-step and its diff for baby-step. To find the test sensor value anomaly probability, a binary search is carried out on the corresponding $sortedSenList$ (this is the $LSG$ values sorted in ascending order on frequency,  in Algorithm~\ref{alg:extendedTraining}) derived from extended training.  If $testval$ equals the sorted sensor list value, then PrLeft, PrRight  and PrAnom  are computed for this $sortedSenListVal$. 
    
	For nearest $sortedSenListVal1$$<$$testval$$<$$sortedSenListVal2$; on $sortedSenListVal1$,  compute its $PrLeft = 1 - PrRight$ and on $sortedSenListVal2$,  compute its $PrRight = 1 - PrLeft$. I.e., we include the frequency of the tested value in the probability computation. Now that we have the PrLeft and PrRight probabilities  for testval,  1 - PrAnom is computed. 
	The products are then computed for the sliding window lengths and checked against the min, max extended bounds determined in training. An anomaly flag is raised when it is out of bounds. Note that sorting and static probabilities were computed once, stored during training and reused later.

	\section{Experimental Results and Analysis}
	\label{sec:experiments}
	The experiments are carried out on a Windows 11 machine, with an Intel(R) Core(TM) i7-1195G7 @ 2.90GHz processor,  single core (of 4-total) was used for all experimentation presented. The GiBy-core detector (Section~\ref{sec:solution}) and extended algorithm (Section~\ref{sec:extended}) are used. The SWaT normal dataset is used for training and attack dataset is for testing.
	TABLE~\ref{table:att} shows the attacks that are detected by GiBy.
	Attacks 5, 9, 12, 15 and 18 in TABLE~\ref{table:att} had no physical impact on the system. Hence, they are not displayed.
	For Attack 3, the 1 mm rate of change is within the range of normal behaviour of the tested state. Hence, this rate of change is not flagged for one of two attack states. The extended algorithm can detect the anomaly when the sliding window (multiplied) probability exceeds the seen threshold in training.
	Attack 13, 14, and 17 are not detected because the changing MV-304 and MV-303 states all showed normal DPIT-301 readings in attack dataset.
	Attack 24 on P-203 is not detected because despite turning on this pump, the pH sensor readings were in the normal range seen in training. The same is also true with Attack 29 on P-203.

	The time taken to train and test the normal and attack SWaT datasets are shown in TABLE~\ref{table:timetakentrain} and TABLE~\ref{table:timetakentest}, respectively. The total training time per sensor was seen to be 9 to 12 minutes. For testing, the total time to copy the nn-actuator(s) and sensor to a new spreadsheet, and computing diff (if any) in the attack dataset are determined in pre-testing steps. The total pre-test time shown is for a total of 449919 attack dataset records. 
	The test step (includes detecting anomaly and providing explanation) is typically seen to be under 1 millisecond per sensor.
	The time taken for training and pre-testing are linear to the number of records seen in the normal and attack datasets, respectively.
	
	For ECOD and Deep-SVDD in Mathuros et al.~\cite{Mathuros2024}, the training time is shorter in comparison with GiBy. This is because they train approximately 24 minutes of data, i.e., $24\cdot 60 = 1440$ records for each of the six process stages of SWaT; as opposed to the entire normal dataset of 495000 records trained by GiBy. ECOD detection time was reported to range from 0.72s to 4.62s, whereas for Deep-SVDD, it was 15.88s to 16.45s. Their Integrated Gradient (IG) XAI used with Deep-SVDD was the fastest with a time of 5.33s. However, the XAI provided there ranks the features in non-increasing order of likelihood; to explain what feature most likely caused the anomaly. With GiBy, the explanation is precise and pinpoints to the sensor and nn-actuation state (see line 6 in Algorithm~\ref{alg:testing}).

	\subsubsection{Analysis}
	\label{ssec:experimentdiscussion}

	The attacks in TABLE~\ref{table:att} --- 3, 13, 14, 17, 24, 26, 29 were either undetected or partially detected. This is because the range of sensor values in the attack dataset were also seen in the training/normal dataset. However, they are considered as attacks because the sensor readings changed even though it remained below the detection threshold. 
	For example, in attack 29, P-203 is ON and HCl flowed out of the chemical tank and was mixed with water. However, the pH sensor AIT-203 readings remained at levels seen in normal dataset training, though it resulted in wastage of chemical (HCl in this case).
	It is possible to detect other edge case attack types such as attack 3 and 6; where the sensor reading is either a fixed value or changes by a constant value but remained below the detection threshold.
	Such a detection may be carried out by observing the number of times the sensor reading pattern repeats and then assigning probabilities for that event.
	Also, for any detector training there is an entropy loss; where the training dataset input is compressed into a smaller trained output (for example see Fig.~\ref{fig:evolution}).
	For these reasons, we avoid trying to detect specific attacks and designed GiBy to be a generalized one-class classifier. However, the implementer is free to plug-in their edge-case detection models onto GiBy classifier.
	Without any plug-in, GiBy in TABLE~\ref{table:att} has 33 Y's and 11 N's in the Detected column. This translates to a 75\% detection of the attacks in the table. 
	Since attack 4 and attack 29 on P-201 has non-existent training data, a correct representation of detected attacks is 33/42, which is 78.5\%.
	We emphasize that the detection accuracy of 78.5\% is specific to attacks in TABLE~\ref{table:att}. 
	An adjusted detection accuracy presented in the abstract is explained and justified in Section~\ref{ssec:operatorworkload}.
	Next, we use an example attack on SWaT to show the fallibility of all anomaly detectors against subtle sensor reading manipulation.
	
	\subsubsection{Undetectable attacks}
	\label{ssec:undetectable}
	Despite best efforts, a powerful attacker who is able to compromise the PLC may introduce a small constant drift $
	\delta$ to the sensor reading to avoid detection.
	Consider the following attack scenario where an attacker has successfully compromised the PLC. Let  an attacker manipulate the level sensor LIT-101 reading with the goal to underflow the tank and dry pump using P-101 (see Fig.~\ref{fig:swat_stage1}).
	Further, let the lower limit for \underline{LIT-101 be 10 mm}; below which the PLC turns OFF the pump P-101; to prevent dry pumping.
	For the tank filling state MV101 = OPEN and P101 = OFF, the attacker would spoof the LIT101 value on the PLC to $Levelval_i = Levelval_i + (\delta \cdot i)$; at each time step $i=\{1,2,\cdots, tmax\} \in tsteps$. This causes the water level to appear to be higher than it actually is.
	For the tank draining state MV101 = CLOSE and P101 = ON, the attacker would spoof the LIT101 value on the PLC to  $Levelval_j = Levelval_j +\delta\cdot (i+j)$; again, causing the tank to appear fuller than it is. By adjusting the value of $\delta$  and the number of time-steps $i$, it results in tank under-flow and P-101 dry pumping. For example, consider LIT-101 value is 12 mm. Let $\delta=0.01$ and $tsteps = 200$.  When tank is in filling state for 150 time-steps, $Levelval_{150} = Levelval_{150} +(0.01\cdot 150)$. Our accumulated drift is now 1.5 mm. Next, let the tank be in draining state for the next 50-time steps. Then, $Levelval_{50} = Levelval_{50} + 0.01\cdot (150+50)$. This causes additional drift of 0.5, and a total drift of $1.5+0.5=2$ mm. As the tank continues to drain, and LIT-101 sensor reading arrives at 11.9 mm, in reality the tank level is at 9.9 mm. The PLC control logic considers the tank level to be above the set lower-bound, but tank underflow occurred, and P-101 is dry pumping.  To make detection further intractable, the drift $\delta$ may be further reduced and the time-steps increased. Further, this attack may be propagated on one or more sensors simultaneously and might be used to create synchronized failures when additional system information is available.     
	Such subtle attacks tailored to remain below the detection threshold are difficult to detect (refer to Section~\ref{sec:conclusions} for possible remedial steps).

	\section{Discussion}
	\label{sec:discussion}

	\subsection{Detector limitations}
	\label{ssec:limitations}
	
	Our core detection solution (see Section~\ref{sec:solution}) is limited to its nn-actuators, for any given sensor.
	This is not necessarily true when the sensor-actuators(s) relationship are available from the system design document, or when a system expert is able to deduce those relationships. In our work, we assumed such additional information is unavailable.
	
	We mapped a smaller state-space of actuator states to the sensor reading. As a result, we forsook the determination of a larger number of tighter individual bounds that takes exponential time for the convenience of a practically smaller solution seek space.
	However, our sensor measurements are taken directly from the normal operations dataset. As a result, the overall determined [LB, UB] bounds remained unchanged, and hence does not change the overall safety limits determined.
	For example, consider eight dependent actuators influencing a sensor reading. When the actuation sequence for each actuator is $k=\{$on, off$\}$, then $|k|=2$. 
	When $|k|=2$ and number of actuators is 8, there are a total of $2^8=256$ actuation states.
	This requires determination of $[LB_i, UB_i]_{\forall i=1,\cdots,256}$ bounds. Additionally, let us assume that only 3 out of the 8 dependent actuators are nearest neighbours, w.r.t. sensor $s$. Using GiBy, we would have only considered nearest neighbour ($2^3=8$) actuation states, and determined $[LB_j, UB_j]_{\forall j=1,\cdots,8}$. 
	As a result, there was no change in the overall state sensor bounds (safety limits) determined, although some detection resolution was lost by considering only the nearest neighbours. 
	
	Another limitation expected from the anomaly detector is a sensitivity to out of bound measurements, resulting in warnings being issued when no attack is present. This limitation could occur even in the system normal (initial) operating phase.
	The normal dataset is expected to cover its everyday boundaries of operation. 
	However, in some cases the normal dataset may not cover all boundaries of normal operations, and these normal boundaries may be surpassed during future plant operations.
	Any logged warning observed, that is deemed as normal, will have to be reviewed and trained on, to avoid those warnings from re-appearing. However, this will stabilize once sufficient normal (training) data indicative of its range of operations is available. This is a training dataset limitation that affects our model.
    	It is clear from Fig.~\ref{fig:nonlinear}, that GiBy does not maintain a global state of time-series events. All detection events are local to the nearest neighbour(s) of the sensor. While this works in water treatment systems,  other systems may additionally require the detection algorithm to keep track of the global state of certain events to detect attacks that work by altering the timeline of events. 
	A final limitation is that it may not detect all invalid actuation states. It is limited to the resolution of the actuator states seen for the nn-actuators. Caution must be exercised in systems where actuators that are not nearest neighbours; whose actuation results in unsafe operations ---  may not be detected by GiBy using only nn-actuators. The training will need to include those other actuators that are in relationship with the sensor.  
	
	\subsection{Detection in an early prototype system design}
	\label{ssec:designlimit}
	In Fig.~\ref{fig:swat}, we note that hydrochloric acid (HCl)  is added into the first tank. 
	Water is pumped using P101 and combined with the HCl pumped by P201. The pH is then measured using sensor AIT201. However, if the HCl being transported is mislabelled as NaOCl, each is poured into an incorrect tank. This means the tank holding HCl now holds NaOCl and vice versa.  AIT201 no longer measure the pH of water mixed with HCl, and it instead measures the mix with NaOCl.
	Now, if the dosing rates are different, the chemicals will be administered at an incorrect rate. For this scenario, it is not possible to detect the pH of water after the addition of HCl, without the assistance of a redundant pH sensor downstream. This design shortcoming is rectified in the actual SWaT testbed. The normal dataset provided is indicative that NaCl was used in the first tank, HCl in second and NaOCl in the third tank. This is because the sensor measurements from the dataset, corresponded with conductivity (AIT201), pH (AIT202) and oxidation reduction potential (AIT203).
	An inadequate design may lead to detection failures, for no fault of the detector modelled.
	We emphasize that good detection capabilities are reliant on   meticulous system design.  
	
	\subsection{GiBy Detection capabilities}
	\label{ssec:capabilities}

	In the threat model (Section~\ref{ssec:threatmodel}), the attacks are classified into single-stage single/multiple component attacks and multi-stage single/multiple component attacks.
	The component considered is a sensor or an actuator.
	The goal of this classification is to build a quantifiable measure (score) on the quality of the detector, w.r.t. its ability to detect these types of attacks. 
	Without loss of generality, the classification may be merged into single and multiple sensor/actuator attack(s) on a system. 
	
	Our core detector (see Section~\ref{sec:solution}) can detect single sensor or actuator manipulations, when it breaks the safety bounds of the relation between sensor and its actuation state.
	When both sensor and corresponding actuator(s) are spoofed, it is no longer able to individually detect that attack, but it depends on whether any of the sensor readings downstream\footnote{Downstream  bounds are also captured by GiBy, since it is computed for every sensor with nn-neighbour actuator(s).}, as a result, went out of bounds. For example, in Fig.~\ref{fig:swat}, consider an attack where P101 is spoofed to ON and FIT201 reading to $normalFlowRate$, despite the pump being OFF and no water flowing. Assume the controller logic is to turn on the first HCl dosing pump (P201), when the FIT201 flow rate $y$ is between $0<x<y<z$ and to OFF otherwise.
	The attack causes the dosing pump to release HCl even when no water is pumped\footnote{ The success of this attack depends on whether the attacker can compromise  the controller PLC and manipulate its measurements.}. However, due to pH sensor AIT201, this attack is detected. Further, if we assume that AIT201 is also spoofed to show normal range of measurements --- a redundant downstream sensor AIT501 incorporated in $P5$ is able to detect the abnormality in pH.
	However, if AIT501 is also spoofed, it may not be able to detect the attack from the measurements made.
	As a side effect, changing the pH might possibly change its conductivity sensor reading, permitting the attack to be detected. It would be difficult to conclude without further experimentation.
	In some scenarios, it may be sufficient to detect the anomaly at a later stage, and in others, it requires to be detected in the same stage. For example, if pump P101 is running above its safe rpm, it might cause the adjacent (following) pipe to burst, affecting the plant and personnel. Where such a concern exists, the design has to include a pressure sensor. When there is a risk the pressure sensor reading is spoofed, a physical pressure safety (release) valve will have to be incorporated into the design~\cite{ISO2025}. An anomaly detector’s ability to detect single/multiple attacks is highly dependent on intrinsic design relationships and system dependencies. This tight coupling makes generalized interpretations on detection capability (by only considering the detection model) less meaningful. 
	
	\subsection{Non-neighbor PLC imposed actuation}
	\label{ssec:plcimposed}
	
	The  $P_2$ stage in Fig.~\ref{fig:swat} involves chemical dosing. Dosing tanks are required to be in the order NaCl, HCl and NaOCl (see Section~\ref{ssec:designlimit} for reason). AIT201 in the SWaT dataset measures conductivity. According to the PLC control logic sheet, the conductivity sensor AIT503 (downstream sensor in $P_5$) read by the PLC instructs P201 to be turned off, when this sensor value hits a preset threshold.
	Turning off the salt pump, in-turn regulates the conductivity sensor reading at AIT503.
	Hence, P201 is additionally considered as a nn-actuator (though not a neighbour) for the sensor AIT503; to capture this relationship. It is added to train the sensor (using GiBy). In general, the PLC logic sheet is examined to include such ‘additional'  nn-actuators, when there is a sensor dependency.
	Not including this relationship does not change the overall detection bounds captured in training. However, it means that some detection resolution is lost because we are no longer directly capturing the relationship between AIT503 and pump P201.

	\subsection{Explainablity}
	\label{ssec:Explainablity}
	
	Anomaly detection is typically followed by a mitigation response, when required. However, extra processing is required to identify why the detector raised a warning; to inform the plant operator of the detector reasoning. This is to assist with the next steps such as diagnostics, emergency manual shutdown, or plant recovery. For example, the anomaly detector may flag time index $t=34323^{th}$ second from start of operation as anomalous. This time index has 51 sensor and actuator readings (see TABLE.~\ref{table:device}). Without further explanation, it is not possible to deduce \textit{i}.) what sensor(s) readings detected the anomalies, \textit{ii.}) what change in actuation states led it to an anomalous state, and \textit{iii.}) what bound it breached.
	
	The more sophisticated AI/ML models using neural network and deep learning are inherently black box models. They are used to detect anomalies but require additional support from eXplaianable AI (XAI) models to explain the decision made by the AI detector. However, both the ML and XAI models have individual accuracy limitations and this is compounded when used in combination. It also takes longer to train and test these models (see Section~\ref{sec:experiments}).
	On the contrary, explainability with the proposed GiBy detector is straightforward. Note that we employed a switchboard in Section~\ref{ssec:switchboard}. Traceability is offered via the one-to-one and onto map, using the switchboard ---  also in the inverse direction, from sensor reading (under testing) to its  [LB, UB] bounds (determined in the training phase).  
	Our training and testing are carried out per sensor, for all concerned sensors and nn-actuator(s). When an anomaly is detected, this immediately tells us what sensor readings are causing anomalies. Since the sensor is known, we can look at its current and previous actuation states, from the state log.
	GiBy raises an anomaly warning when \textit{i}.) the sensor reading (under testing) for the nn-actuator state is out of bound. Our detector is able to provide an explanation remark, whether it was because the sensor reading was $>UB$ or $<LB$. \textit{ii}.) when the normal training dataset did not have the nn-state actuator state for the sensor, in it. This may happen for two reasons \textit{a.)} the actuation state is invalid or \textit{b.}) that state is yet to be encountered in the plants' normal operation and hence not yet seen in the normal training dataset.
	Our detector system is designed to offer a level of explainability that is expected to equip the operator to deal with most contingencies. The capability of our system to provide the correct sensor location and response to faulty components/attacks subsequently reduces the plant downtime.
	
	\subsection{Operator Workload Reduction}
	\label{ssec:operatorworkload}
	
	We revisit the definitions of   precision, recall, accuracy, and F1-score in Equations~(\ref{eqn:precision})-(\ref{eqn:f1score}), and discuss what it means for the operator.
	\begin{equation}
		\label{eqn:precision}
		Precision = \frac{True Positive (TP)}{True Positive (TP)+False Positive (FP)}
	\end{equation}
	In our case, GiBy is an one-class classifier and it considers all sensor readings within the determined sensor boundaries as safe.
	For the purpose of reducing the operator workload, we define (for this section) any sensor reading within the safety boundary to be non-anomalous.
	This implies that even if a sensor reading had changed but its resulting value was within the safety limits; it is declared a safe and non-anomalous operation.
	A false positive is recorded when the model incorrectly predicts a positive outcome (within safety boundary) as an outcome that is  negative (outside safety boundary). For example, a false positive occurs when a safe sensor reading value is predicted as unsafe (anomalous). Training is carried out on a normal dataset to determine safety boundaries. The safety boundaries  expand when further training data is made available; and eventually the rate of expansion stabilizes --- once it has seen most of the normal operations boundaries.  
	A false positive limit may be chosen by the operator and GiBy may be used once it meets this requirement. This allows an operator to manage the number of FPs seen.

	\begin{equation}
		\label{eqn:recall}
		Recall = \frac{True Positive (TP)}{True Positive (TP)+False Negative (FN)}
	\end{equation}
	
	A false negative is recorded when the model incorrectly predicts a negative outcome (outside safety boundary) as an outcome that is  positive (within safety boundary).
	In GiBy, there are no FNs because as it continues to train on normal data, the safety boundary expands outwards from a point. 
	
	Since GiBy has no FN and its FP is determined by the length and breadth of the boundaries seen in normal training data --- we argue that GiBy's true positive and true negative are a measure of its detection capabilities. Note that in Equation~(\ref{eqn:precision}), as $FP\rightarrow$0, $Precision\rightarrow$1. In Equation~(\ref{eqn:recall}), as $FN\rightarrow$0, $Recall\rightarrow$1. For GiBy, as more training data is available, its  $FP\rightarrow$0 and $Precision\rightarrow$1. Since there are no FNs in GiBy, its $Recall=1$. This implies, an extensively trained GiBy that has seen most boundaries will have $Precision\approx 1$ and $Recall=$1, resulting in a F1-score close to 1; w.r.t. the TP, FP and FN the operator sees.

	\begin{equation}
		\label{eqn:accuracy}
		Accuracy =  \frac{TP+TN}{TP+TN+FP+FN}
	\end{equation}

	\begin{equation}
		\label{eqn:f1score}
		F1-score = 2\cdot \frac{Precision\cdot Recall}{Precision+Recall}
	\end{equation}
	
	When FP $\rightarrow 0$ (as is the case for it to be practically useful for the operator), its precision is also dominated by the TP.
	The operator is now tasked with resolving actual anomalies and the occasional FP. The FP frequency of occurrence is loosely bounded by the false positive limit (threshold). This is a configurable parameter set by the implementer during setup. However, it is required to meet the average FP frequency threshold set for usability.
	
	\subsubsection{GiBy Re-defined Accuracy}
	\label{sssec:redif}
	When we used the re-definition for accuracy shown in Equation~(\ref{eqn:accuracy}) --- it additionally considers any anomalous (changed) reading within the safe operation limits as not anomalous, to reduce the operator workload.
	Further, it is straight-forward to report any state not seen in training as anomalous/not seen. 
Now, with reference  to TABLE~\ref{table:att}, attacks 13, 14, 17, 24, 27, and 29, will be considered as safe and non-anomalous. Only attack 3 is undetectable by GiBy. The re-defined accuracy is  43/44 = 0.9772 or 97.72\%. We maintain that the detection accuracy of 78.5\% (based on a conventional\footnote{For example, consider an email spam filter model, that correctly classifies 90 out of 100 emails. Its detection accuracy is 90/100=0.9 = 90\%. Even for a single email with 1000 detected spam words, it is counted as one spam email, unless  a contextual reasoning for re-defining accuracy is provided.} definition~\cite{Anna2016}) reported in Section~\ref{ssec:experimentdiscussion} is high. This is because GiBy correctly detected all anomalies outside its normal (seen) trained bounds.

	\subsection{Non-standard detection accuracy reporting}
	\label{ssec:xfiles}
	In this section we highlight certain techniques used to increase detection accuracy by  changing the definition for accuracy seen in Equation~(\ref{eqn:accuracy}).  
    The purpose of this discussion is to highlight the different definitions for the seemingly same reference to the word ``detection  accuracy'', and why some works in existing literature are not comparable with GiBy's detection accuracy.
	The interpretation of what accuracy measures and how it is reported is inconsistently defined and calculated; making GiBy detection accuracy comparisons  incompatible with some works in existing literature.
	The work in Inoue et al.~\cite{jun2017} reports using SWaT dataset for their DNN model, a precision of 0.98295. For One-class SVM, their reported precision is 0.92500. It noted that false and true positive rates for DNN were counted over log entries and over windows for SVM.
	In GiBy, the accuracy is determined by Equation~(\ref{eqn:accuracy}) (also see Section~\ref{ssec:experimentdiscussion} and Section~\ref{sssec:redif}). This is incompatible with the definition of log entries and windows used.
	The work in Xu et al.~\cite{Xu2024} reports SWaT detection accuracy  for different models ranging from 95.97\% to 99.95\% (refer to their attack classification performance on the SWaT Dataset table). They use a  downsampled SWaT dataset contains 47044 samples. This included 26877 normal dataset samples. However, for GiBy --- we use the full range of values in the training and testing dataset discussed in Section~\ref{ssec:swat}.
	Further, there are a total of 41 attacks on SWaT shown in TABLE~\ref{table:att}. Even if 40 out of 41 attacks were detected, the accuracy would be $40/41=0.9756$. To arrive at an accuracy of 0.99, the division has to be no less than 99/100 = 0.99. In Xu et al.~\cite{Xu2024}, they report accuracies up to 0.9995, which is a mathematical impossibility; for how accuracy is defined for GiBy (see Section~\ref{ssec:experimentdiscussion} and Section~\ref{sssec:redif}). This clearly indicates a different definition was used for their detection accuracy;  making any comparisons with GiBy incompatible.

	\subsection{A real-time and scalable solution}
	\label{ssec:scalable}
    The GiBy algorithm presented in Section~\ref{sec:solution} is universally transferable and GiBy-core testing may be parallelized. This is because it produces a series of independent [LB, UB] boundaries.
    The computational cost of GiBy-core tests (see Section~\ref{ssec:giby}) are to see whether the sensor value is in between two bounds [LB, UB]; taking up to two comparisons. The main costs of the extended algorithm tests (see Section~\ref{sec:extended}) are the binary search and sliding window multiplications in each time step. Its time complexity may be reduced to that of binary search on the pre-sorted linearized group (that takes logarithmic time), and a maximum of one multiplication and division: for each time step. Instead of repeatedly multiplying every sensor derived value in the sliding window, the newest time step sensor derived value may be multiplied into the existing product, and the sensor derived value at the tail of the window is divided out.
    From TABLE~\ref{table:timetakentest}, it is seen to take under one-millisecond of processing time to determine whether a sensor reading is anomalous or not. This processing time includes the giant-step (and/or) baby-step testing and explainability presented in Section~\ref{sec:solution}, and the extension detection testing presented in Section~\ref{sec:extended} --- for the experimental setup in Section~\ref{sec:experiments}. 
    
    GiBy is also highly scalable. This is because all the testing are carried out per sensor. Its time complexity is bounded by $\mathcal{O}(n)$; $n$ being the total number of sensors tested for anomaly. Note the experimental setup in Section~\ref{sec:experiments} employs only one  of four available computational cores, for the test results achieved in TABLE~\ref{table:timetakentest}. Hence, they are shown to be practical real-time solutions to detect anomalies in water treatment plants.

	\subsection{Usability in other sectors}
	\label{ssec:Usability}
	It is clear from Fig.~\ref{fig:nonlinear}, that GiBy does not maintain a global state of time-series events. All detection events are local to the nearest neighbour(s) of the sensor. However, in some systems, it may additionally require the algorithm to keep track of the global state of certain events to detect anomalies. 
	Anomaly detection and mitigation in some sectors (such as power), in certain scenarios, require an actuation within 10-15 milliseconds from event occurrence, to ensure safety.  It may be pre-programmed into the control logic of an Intelligent Electronic Device (IED), typically used in a power substation. It may be used to trip the circuit breaker in near real-time (when sensor readings are out of safety bounds), or to inform an operator regarding the anomalous behaviour.
	Due to the straightforward mappings and simple boundary checks in GiBy (see Section~\ref{ssec:scalable}), it might also find (outlier detection) applications in resource constrained devices, edge computing devices and legacy systems with lower computational power.

	\section{Related Work}
	\label{sec:litreview}
	
	The focus of this work is on process anomaly detection~\cite{Raman2022, Saheed2023}.
	Several approaches are available for design-centric and data-centric anomaly detection.
	In a design-centric approach, for example, Adepu and Mathur~\cite{Adepu2016DDSSMP} captured the design using state condition graphs. It was used to capture the conditions on actuation w.r.t. sensor readings, expressed using Boolean conditions, and represented in a graphical format. 
	The work in Yoong et al.~\cite{Yoong2021}  used axiomatic design methodology for systems, where it iteratively decomposes a CPS design to sets of dependent components and transformed into invariants.
	The work in Raman and Mathur~\cite{Gauthama2022} has a detection rate of over 90\% against actuators and above 75\% against sensors, excluding certain stealth attacks. In GiBy, out of bound sensor values are considered as anomalies; irrespective of whether they were indirectly caused by changes in actuator state or directly by altering the sensor value.
	In Merwa et al.~\cite{Merwa2024}, an association rule mining technique was used to generate attack and invariant rules. 
	In a data-centric approach, for example, there exist solutions to distinguish outliers, such as ECOD in Li et al.~\cite{ECOD}.
	It computed a univariate empirical cumulative distribution function for each dimension separately.
	To measure the chance of a data point being anomalous, they computed its tail probability across all dimensions.
	Another outlier detector called deep Support Vector Data Description (Deep-SVDD) proposed in Ruff et al.~\cite{SVDD}, trained a neural network
	while minimizing the volume of a hypersphere that enclosed
	the network representations of the data.
	In WaXAI,  Mathuros et al.~\cite{Mathuros2024} employed ECOD and Deep-SVDD in the context of SWaT anomaly detection. In addition, they also employed XAI~\cite{Ali2023} models, namely,  kernel SHAP~\cite{SHAP}, LIME~\cite{LIME}, ALE~\cite{ALE} and IG~\cite{IG}, to provide explanation for the anomaly detected.
	Further, a 20\% improvement in attribution of attack root cause over SHAP was achieved by using a Factorization Machines based approach by Avdalovic et al.~\cite{avdalovic2025}. 
	In this work we provided attribution (explainability) linked directly to the sensor. Explainability was provided for each sensor when out of bounds, to top up the recommendations in Fung et al~\cite{Fung2024}.
	GiBy showed millisecond detection speeds on an everyday used microprocessor demonstrating that the algorithm is energy efficient.
	
	\section{Conclusions and Future Work}
	\label{sec:conclusions}
	We presented a simple yet powerful anomaly detector that trained four set of bounds --- the min-max bounds for the giant-step, the rate of change bounds for baby-step, and the extended detection algorithm bounds, for both giant and baby-steps. The detector acted as a one-class classifier. Any tested value outside trained bounds were flagged as an anomaly.
	The transparent design of GiBy anomaly detector made it straightforward to implement and easy to interpret. The explainability provided pinpoints the sensor and actuation state for which anomaly was detected, and what bounds it breached. 
	The experiments showed that anomaly detection and explanation took around
	$\dfrac{1}{1000^{th}}$ of a second per sensor (on average).
	This makes it useful for implementation in systems where near real-time decisions are made, or on devices that are resource constrained. 
	Due to the undetectable attacks example discussed in Section~\ref{ssec:undetectable}, in future work we will explore other types of detection using  digital signature verification and encryption to hide the inferences on sensor readings across OT, segregating and air-gapping high risk sensors and PLCs, paying closer attention to attack prevention and fast recovery strategies that do not compromise plant and personnel safety in the meantime. 
	
	\small 
	\section*{Acknowledgment}
	This research is funded by UK Research and Innovation, Knowledge Transfer Partnership project, KTP reference 13504. The authors are grateful for the discussions with Stephen and George, leading to the manuscript.

\bibliographystyle{elsarticle-num}
\bibliography{biblio}


\end{document}